\documentclass[twocolumn]{aastex631}

\usepackage{amsmath, amssymb}
\usepackage{float} 
\usepackage{threeparttable}
\usepackage{microtype}
\usepackage{enumerate}

\usepackage{etoolbox}  

\newcommand{\cMpc}{{\rm cMpc}}
\newcommand{\kpc}{{\rm kpc}}
\newcommand{\ckpc}{{\rm ckpc}}

\newcommand{\cm}{{\rm cm}}

\newcommand{\K}{{\rm K}}
\newcommand{\Tvir}{{$T_{\rm vir}$}}

\usepackage{xcolor}

\def\kms{\ifmmode {\rm km\:s^{-1}} \else $\rm km\:s^{-1}$\fi}

\usepackage{graphicx}

\begin{document}
\title{Multiphase Gas in High-$z$ Cosmic Sheets}

\author[0000-0001-5822-5622]{Charles G. Willard}
\affiliation{University of Washington | Seattle, WA, USA}

\author[0000-0003-3520-6503]{Yakov Faerman}
\affiliation{University of Washington | Seattle, WA, USA}
\affiliation{School of Physics and Astronomy, Tel Aviv University, Ramat Aviv 69978, Israel}

\author[0009-0008-5043-6220]{Samuel McCarty}
\affiliation{University of Washington | Seattle, WA, USA}
\affiliation{Center for Astrophysics | Harvard \& Smithsonian}

\author[0000-0001-7961-9735]{Matthew McQuinn}
\affiliation{University of Washington | Seattle, WA, USA}

\author[0000-0001-8057-5880]{Nir Mandelker}
\affiliation{Centre for Astrophysics and Planetary Science, Racah Institute of Physics, The Hebrew University, Jerusalem 91904, Israel}
\affiliation{University of Washington | Seattle, WA, USA}

\author[0000-0001-5510-2803]{Thomas Quinn}
\affiliation{University of Washington | Seattle, WA, USA}

\begin{abstract}

We present high-resolution cosmological simulations of individual cosmic sheets ($z=3-6$) using the moving-mesh code AREPO to study cold gas ($T\approx10^4\rm ~K$) in the diffuse cosmic web. We find cosmic sheet gas forms a fragmented morphology, evident as kiloparsec-scale variations in the neutral hydrogen column density ($N_{\rm HI}$). Fragmentation arises from the bending and disruption of the cold-dense sheet core into a multiphase medium. Nonlinear thin shell instability (NTSI) and later shear-induced vorticity drive this bending, rather than previously suggested thermal or Richtmyer-Meshkov instabilities. Higher resolution enhances fragmentation. Across a fourfold range in spatial resolution, $N_{\rm HI}$ covering fractions increase by factors of $\approx 1.5-3.5$ for partial Lyman-limit systems (pLLS) and Lyman-limit systems (LLS). An order-of-magnitude estimate suggests sheets may host a large fraction of the ultra-low metallicity (p)LLS population. Equivalent MP-Gadget simulations show qualitatively similar fragmentation, but numerical smoothing effects are present. Cosmic-sheet gas properties remain unconverged despite our high resolutions.
\end{abstract}

\section{Introduction}

Matter in the universe is distributed in a ``cosmic web'' of sheets, filaments, and nodes, which form through anisotropic gravitational collapse of primordial overdensities in one, two, and three spatial dimensions, respectively \citep{Zeldovich1970, Bond1996, Springel2006}. Most cosmic baryons reside in the intergalactic medium (IGM), primarily in sheets and filaments \citep{Rauch1998, Cautun2014, McQuinn2016, Martizzi2019}. The density and temperature of IGM gas evolve through gravitational compression, photo-heating, shock-heating, and radiative cooling as it assembles into the cosmic web \citep{McQuinn2016}. 

While much work over the last several years has focused on intergalactic filaments, hardly any attention has been paid to cosmic sheets. Sheets are geometrically the simplest elements of the cosmic web, arising primarily from collapse along a single dimension and, at high redshifts, are unaffected by poorly understood galactic feedback. Yet their baryonic properties remain poorly understood. In this work, we explore the gas properties and dynamics of cosmic sheets. 

Early work on sheets, analytic and 1D simulations, established their collapse physics from primordial density perturbations and their rough baryonic structure: a cold-dense core bounded by hot virial layers with an accretion shock \citep{Zeldovich1970, Sunyaev1972, Jones1981, Shapiro1983, Bond1984, Shapiro1985}. 2D simulations of individual sheets demonstrated that sheets could be prone to gravitational and hydrodynamic instabilities \citep{Anninos1994, Anninos1995, Valinia1997}.

In later 3D cosmological simulations, work shifted from individual sheets to the cosmic web \citep{Cen1993, Cen1994}, connecting sheet and filament populations to primordial overdensities as well as to high-$z$ Lyman-alpha absorption \citep{Cen1994, Bond1996, MiraldaEscude1996}. Today, cosmic sheets are most often studied at this global level, using algorithms to systematically classify simulation volumes into voids, sheets, filaments, and nodes \citep{Hahn2007, AragonCalvo2007, ForeroRomero2009, Cautun2014, Martizzi2019}. 

More recently, \citet{Mandelker2019} performed the first modern cosmological zoom simulation focused on an entire high-$z$ cosmic sheet (IPMsim, IPM stands for intrapancake medium). The fine spatial resolution of IPMsim enabled (sub)kpc-scale studies of cosmic sheet gas, cosmic filaments, and the environmental effects of sheet/filament virial gas on dwarf galaxies \citep{Mandelker2021, Lu2023, Pasha2023}. Most relevant here, IPMsim revealed a complex morphology in the cold phase sheet gas at high resolution; we will call this phenomenon fragmentation. Fragmentation yields strong spatial variations in neutral hydrogen column density, potentially observable in quasar absorption spectra.

Diffuse intergalactic sheet and filament gas is most commonly observed via intensity dips in the spectra of background quasars, an absorption signal caused by the Lyman-$\alpha$ (1216 \text{\AA}) resonance of neutral hydrogen \citep{GunnPeterson1965, Meiksin2009, McQuinn2016}. The redshifting of this line along the line of sight (LOS) creates a picket fence of absorption features across a wide range of wavelengths, some of which are from intersecting cosmic sheets; this is the Lyman-$\alpha$ forest (LyaF). The LyaF traces neutral hydrogen column densities of $N_{\rm HI}<10^{16} ~\cm^{-2}$, which are optically thin to the hydrogen-ionizing continuum. Absorbers in the transition region from optically thin to thick absorption are called partial Lyman-limit systems (pLLS) with $10^{16.2}<N_{\rm HI}<10^{17.2} ~\cm^{-2}$, whereas Lyman-limit systems (LLS) are optically thick at the Lyman-limit, $10^{17.2}<N_{\rm HI}<10^{19}~\rm cm^{-2}$. 

The numerical convergence of high-redshift cosmic sheet gas in cosmological simulations remains unclear. One might expect cosmic sheets to be well converged in simulations given the $\approx1\%$ convergence of LyaF flux statistics and the $\approx10\%$ convergence of the column-density distribution, at $10^{12.5}<N_{\rm HI}<10^{15}~\rm cm^{-2}$ \citep{Lukic2014}. For higher columns, $N_{\rm HI}>10^{17}~\cm^{-2}$, \citet{Rahmati2013} likewise found $\mathcal{O}(10\%)$ convergence for the column-density distribution. However, IPMsim found a threefold higher LLS covering fractions across a fourfold increase in spatial resolution \citep{Mandelker2019, Mandelker2021}, suggesting cosmic sheet gas is poorly converged.


To revisit the topic of cosmic sheet gas properties and morphology on small scales, we carry out a suite of idealized cosmic sheet collapse simulations across a range of resolutions. Going beyond previous work, we study how fragmentation manifests in sheets of varying virial temperature, examine the role of small-scale perturbations in seeding fragmentation, and identify the driving mechanism(s). Section~\ref{sec:Methods} details our simulation suite. Section~\ref{sec:IGMGasFragmentation} presents cosmic sheet fragmentation in our fiducial simulations. Section~\ref{sec:NumericalTesting} studies fragmentation in modified sheet simulations, with altered power spectra or collapse histories. Section~\ref{sec:Discussion} discusses the driving mechanism for sheet gas fragmentation, sheet contributions to absorber populations, and the role of numerics for sheets in smoothed particle hydrodynamics. We conclude in Section~\ref{sec:Conclusion}.

\section{Simulation Methods} \label{sec:Methods}

We present a suite of small-box, high-resolution cosmological simulations of individual cosmic sheets. We use the moving-mesh, Voronoi-tessellation cosmological hydrodynamics code AREPO \citep{Weinberger2020}. AREPO evolves gas and dark matter according to gravitational and magnetohydrodynamic (MHD) physics in an expanding universe, including subgrid models for an ionizing UV background, atomic cooling, star formation, and stellar feedback. In Section \ref{subsec:SimSetup}, we present our simulation setup, including initial conditions (ICs) and code parameter choices, and in Section \ref{subsec:ZA_Purterbations}, we detail the method used to seed the collapse of a cosmic sheet in our boxes.

In Appendix \ref{appendix:SPH}, we present the same fiducial simulation suite run using the code MP-Gadget \citep{Feng2018}, which employs smoothed-particle hydrodynamics (SPH). We use AREPO over MP-Gadget in the main text as we found that shock broadening and smoothing over multiphase gas in SPH made separating numerical from physical effects difficult (cf. Section \ref{subsec:Arepo_v_MPgadget}).

\subsection{Simulation Setup} \label{subsec:SimSetup}

Our fiducial suite of simulations consists of two box sizes, $L=\{4, 8\}~{\rm cMpc}/h$. Each box is run at three resolutions, $N_{\rm tot}=2\times\{128^3, 256^3, 512^3\}$ particles, which we will call the low-, medium-, and high-res simulations, respectively. The two box sizes and three particle resolutions yield four effective spatial/mass resolutions. The typical Voronoi-mesh gas cell sizes for dense ($\rho\sim10^{-26}\rm~ g~cm^{-3}$) gas in the cosmic sheet cold cores ($T\sim 10^4~$K) at $z=4$ are $\Delta x_{\rm cell}\approx\{3.2,~ 1.6, ~0.8, ~0.4\}~\rm pkpc$. Simulation parameters, including gas particle masses, are shown in Table \ref{tab:Simulations}.

Our small boxes maximize spatial resolution but suppress structure formation on scales approaching the box size. We therefore force the collapse of a single cosmic sheet by adding a 1D, normalized Gaussian overdensity perturbation along the $x$-axis to the ICs. The perturbation is detailed in Section~\ref{subsec:ZA_Purterbations}. The $L=4$ and $8~\mathrm{cMpc}/h$ simulations produce sheets with virial temperatures of $T_{\rm vir}\approx2\times10^5$ and $8\times10^5~\rm K$, respectively, hereafter the low- and high-$T_{\rm vir}$ sheets. These temperatures were measured from the post-shock gas at $z=4$ from our homogeneous simulations, not derived a priori. The box sizes were chosen to sample a range of cosmic-sheet masses. We estimate there are $\mathcal{O}(10^3)$ low and $\mathcal{O}(10^1)$ high \Tvir~sheets per cosmic volume of $(100~\rm cMpc/h)^3$ at $z=4$ (Appendix \ref{appendix:SheetPops}).

For these cosmic sheets generated by the 1D Gaussian collapse, we consider three different background matter-power-spectrum perturbation scenarios. (1) Fiducial: The full $\Lambda$CDM linear matter power spectrum, allowing filaments and halos to form within the sheet. (2) High-$k$: Only high wavenumber $\Lambda$CDM modes, with $k > 2\pi/(100 ~\ckpc/h)$, suppressing the formation of filaments and halos with $M_{\rm h}\gtrsim10^8~M_{\odot}$ while retaining small-scale seeds for hydrodynamic instabilities, as we will demonstrate. (3) Homogeneous: No density perturbations beyond the imposed 1D sheet perturbation.

Table \ref{tab:Simulations} provides a summary of all 18 of these simulations, plus size additional `modified' test cases that we will discuss in the text. The fiducial set will be the focus of Section~\ref{sec:IGMGasFragmentation}; the high-$k$ perturbation and homogeneous sets will be the focus of Section~\ref{sec:NumericalTesting}.

We use the public version of AREPO \citep{Weinberger2020}. We generate ICs using MP-GenIC \citep{Bird2020}. We use the same initial random seed for all simulations, which means that the phases of the Fourier modes are the same when scaled by the box size. For all simulations we use a gravity softening of 1/30$^{\rm th}$ the mean interparticle spacing. All ICs are generated at a redshift of $z_{\rm init}=99$. We generate the linear matter power spectrum ($P_{m, \Lambda \rm CDM}(k)$) using the Code for Anisotropies in the Microwave Background (CAMB, \citealp{CAMB2000, CAMB2012}) with parameters: $\Omega_m=0.2814, ~\Omega_{\Lambda}=0.7186, ~\Omega_b=0.0464, ~H_0=69.7$.

We adopt a simplified star formation (SF) prescription, without stellar winds, to isolate sheet fragmentation. Star formation is negligible in the inner sheet, far from halos. For computational efficiency, SF is enabled only in the fiducial, full-power-spectrum runs. We use the \citet{Springel2003} SF model with parameters of $\delta_{\rm crit}=1000$, $t_{\rm sfr}=10^{-4}$, $A_{\rm EVP}=1$, $\beta=10^{-4}$ (see parameter definitions in \citealt{Springel2003}, Section 3).

We compared our no-feedback simulations to those using the wind prescription of \citet{Okamoto2010}, with the wind speed, mass loading, and hydrodynamic decoupling parameters used in ASTRID \citep{Bird2022, Ni2023}. Halos and, to a lesser degree, filaments are of course affected by winds, but our analysis focuses on the diffuse cosmic sheet gas and explicitly removes filaments and halos. Winds are negligible for our conclusions regarding cosmic sheet enhancements of pLLSs and LLSs. This testing was originally performed in MP-Gadget, not AREPO, but the inability of feedback to contaminate cosmic sheet gas should hold generally. 

We adopt the ionizing ultraviolet background spectrum of \citet{FaucherGiguere2009} (June 2011 version). The simulations include cooling and heating by collisional excitation, collisional ionization, recombination, bremsstrahlung, and inverse-Compton scattering off CMB photons. We apply the self-shielding correction of \citet{Rahmati2013} in post-processing. We ran one simulation without helium to test whether the shape of the cooling curve played a role in cold sheet gas fragmentation and found it did not, as we discuss briefly in Section \ref{subsec:AlternateMechanisms}. Magnetic fields were not enabled.

\begin{table}
    \centering
    \begin{threeparttable}
        \begin{tabular}{cccc}
             L & $N_{\rm tot}$ & $m_{\rm gas}$ & $P_{\rm bg}(k)$ \\
             $(\rm cMpc/h)$ & & ($\rm M_{\odot}$) & \\
             \hline
             \hline
             \multicolumn{4}{c}{\textit{Fiducial Set}} \\
             \hline
             4 & $2\times128^3$ & $6\times10^5$ & $\Lambda \rm CDM$ \\
             4 & $2\times256^3$ & $7\times10^4$ & $\Lambda \rm CDM$ \\
             4 & $2\times512^3$ & $9\times10^3$ & $\Lambda \rm CDM$ \\
             8 & $2\times128^3$ & $5\times10^6$ & $\Lambda \rm CDM$ \\
             8 & $2\times256^3$ & $6\times10^5$ & $\Lambda \rm CDM$ \\
             8 & $2\times512^3$ & $7\times10^4$ & $\Lambda \rm CDM$ \\
             \hline
             \multicolumn{4}{c}{\textit{High-k Perturbation Set}} \\
             \hline
             4 & $2\times128^3$ & $6\times10^5$ & High-$k$ Only \\
             4 & $2\times256^3$ & $7\times10^4$ & High-$k$ Only \\
             4 & $2\times512^3$ & $9\times10^3$ & High-$k$ Only \\
             8 & $2\times128^3$ & $5\times10^6$ & High-$k$ Only \\
             8 & $2\times256^3$ & $6\times10^5$ & High-$k$ Only \\
             8 & $2\times512^3$ & $7\times10^4$ & High-$k$ Only\tnote{a}  \\
             \hline
             \multicolumn{4}{c}{\textit{Homogeneous Sheet Set}} \\
             \hline
             4 & $2\times128^3$ & $6\times10^5$ & $P_{\rm m}(k)=0$ \\
             4 & $2\times256^3$ & $7\times10^4$ & $P_{\rm m}(k)=0$ \\
             4 & $2\times512^3$ & $9\times10^3$ & $P_{\rm m}(k)=0$ \\
             8 & $2\times128^3$ & $5\times10^6$ & $P_{\rm m}(k)=0$ \\
             8 & $2\times256^3$ & $6\times10^5$ & $P_{\rm m}(k)=0$ \\
             8 & $2\times512^3$ & $7\times10^4$ & $P_{\rm m}(k)=0$ \\
             \hline
             \multicolumn{4}{c}{\textit{Additional Simulations}} \\
             \hline
             6 & $2\times512^3$ & $3\times10^4$ & High-$k$ Only \\
             8 & $2\times512^3$ & $7\times10^4$ & High-$k$ Only\tnote{a}$~^{,}$\tnote{b} \\
             8 & $2\times512^3$ & $7\times10^4$ & High-$k$ Only\tnote{a}$~^{,}$\tnote{c} \\
             4 & $2\times1024^3$ & $1\times10^3$ & $\Lambda \rm CDM$\tnote{d} \\
             8 & $2\times1024^3$ & $9\times10^3$ & $\Lambda \rm CDM$\tnote{d} \\
             8 & $2\times512^3$ & $7\times10^4$ & $\Lambda \rm CDM$\tnote{d, e} \\
        
        \end{tabular}
        \begin{tablenotes}
            \item[a] $z_{\rm end}=3$
            \item[b] Tophat Overdensity Profile, $z_{\rm coll}=6$
            \item[c] Delayed Collapse, $z_{\rm coll}=3$
            \item[d] MP-Gadget Run Only
            \item[e] No Helium
        \end{tablenotes}
    \end{threeparttable}

    \caption{Simulation suite parameters: box size ($L$), total particle number ($N_{\rm tot}$), baryonic particle mass ($m_{\rm gas}$), and background matter power spectrum ($P_{\rm bg}$). The primary suite spans two box sizes, three particle resolutions, and three background perturbation models. Simulations use AREPO, evolved to $z_{\rm end}=4$ unless noted, and include a 1D Gaussian overdensity seed with $z_{\rm coll}=4$. A copy of the fiducial set was run with MP-Gadget. Additional simulations vary the 1D seed geometry or collapse redshift, numerical method, or included physics.}

    \label{tab:Simulations}   
\end{table}

\subsection{Sheet Perturbation} \label{subsec:ZA_Purterbations}

To achieve the collapse of a cosmic sheet, despite the relatively small volumes of our simulations, we seed an idealized 1D overdensity perturbation that collapses at a specific redshift (for a tophat density profile) or redshift range (for a Gaussian density profile). The Zeldovich approximation (ZA) is used to compute the initial linear displacement and velocity field of particles for the 1D perturbation \citep{Zeldovich1970}. Our fiducial scenario uses a normalized Gaussian linear overdensity profile, where linear overdensity is $\delta = \rho/\bar \rho_{M}-1$ and $\bar \rho_{M}$ is the cosmic mean matter density. We also explore one simulation with a tophat overdensity profile in Section \ref{subsec:ModifiedCollapse}, but here we focus on the Gaussian profile. 

A 1D normalized Gaussian overdensity profile at Lagrangian coordinate $\mathbf{q}$ can be expressed as:
\begin{equation}
\delta_{\rm init}(q) = \delta^* \left( \exp \left[-\frac{q^2}{2 q_*^2} \right] - \delta_{\rm norm} \right),
\label{eqn:deltainit}
\end{equation}
where $q$ is the component of $\boldsymbol{q}$ in the direction orthogonal to the sheet ($x$-axis). Subscript `init' denotes the quantity at our initialization redshift of $z_{\rm init}=99$. We set $q_{*}=L/4$, meaning one standard deviation of the Gaussian profile is one quarter of the box size. $\delta_{\rm norm}$ is defined by the constraint that the mean overdensity in the box is zero:
\begin{equation}
\delta_{\rm norm} = \sqrt{2\pi}\frac{q_*}{L}{\rm erf}\left(\frac{L}{2\sqrt{2}q_*}\right).
\label{eqn:delta_norm}
\end{equation}

The comoving displacement to achieve the above overdensity distribution is denoted as $\boldsymbol{\Psi}(\boldsymbol{q},z_{\rm init})$. We will drop the vector notation, as we only need to consider the direction orthogonal to the sheet. The initial displacement is given as
\begin{equation}
    \begin{aligned}
    \Psi_{\rm init} &= -q\bar{\delta}(q), \\
    &= -q \, \delta^* \left[ 
    \sqrt{\frac{\pi}{2}} \frac{q_*}{|q|}\,
    {\rm erf}\!\left(\frac{|q|}{\sqrt{2}q_*}\right)
    - \delta_{\rm norm}
    \right],
    \end{aligned}
    \label{eqn:displacment}
\end{equation}
where $\bar{\delta}(q)$ is the average overdensity interior to the Lagrangian element. The comoving velocity is simply the rate of change of the displacement, $v = \dot \Psi_{\rm init}=H_{\rm init}f_{\rm init}\Psi_{\rm init}$, where, $f=d \ln D/d\ln a$, $a$ is the scale factor and $D(z) = (1+z_{\rm init})/(1+z)$ is the linear growth factor during matter domination normalized to unity at $z_{\rm init}$.\footnote{Our initial redshift of $z_{\rm init}=99$ and collapse redshifts of $z_{\rm coll}=4$ are within the matter-dominated era.} 

The amplitude parameter, $\delta^*$, is solved for once a redshift for sheet collapse ($z_{\rm coll}$) is specified. We define collapse as when the Lagrangian element $q_*$ reaches the origin. Given the ZA mapping, $x(q, z) = q[1-D(z)\bar{\delta}(q)]$, the expression for origin crossing is $D(z_{\rm coll})\bar{\delta}(q_{*})\rightarrow1$.\footnote{One could also choose to define collapse as when the ZA density goes to infinity, $D(z_{\rm coll})\delta(q_{\rm coll})\rightarrow1$; this is where the Jacobian ($dx=J~dq$) vanishes and a caustic forms in ZA theory. For a tophat, the criteria for caustic formation and origin crossing are simultaneous for all elements q in the overdense perturbation. For a Gaussian profile these are distinct conditions. For our purposes, origin-crossing is the more intuitive quantity to define collapse.} Thus we can solve for $\delta^*$ as,

\begin{equation}
\delta^*  = \frac{1+z_{\rm coll}}{1+z_{\rm init}} \left[\sqrt{\frac{\pi}{2}} \mathrm{erf} \left(\frac{1}{\sqrt{2}} \right) - \delta_{\rm norm} \right]^{-1}.
\label{eqn:delta_star}
\end{equation}

Thus, our only model parameters are $z_{\rm init}=99$, $z_{\rm coll}=4$, $q_{*}=L/4$, and $L$. We emphasize that Gaussian collapse to the origin is not simultaneous. The mass elements nearer the center of the Gaussian collapse to the origin before the edges. The formally continued ZA mapping predicts that a mass fraction, $F$, of the total box mass has collapsed to the origin by a redshift $z_{\rm origin}$ as

\begin{equation}
1+z_{\rm origin}(F)
=(1+z_{\rm coll})\frac{\bar{\delta}_{\rm L,init}(FL/2)}{\bar{\delta}_{\rm L,init}(q_*)},\qquad 0<F<1.
\label{eq:zF}
\end{equation}

This gives that $\approx1\%, ~12.5\%, ~25\%$, and $50\%$ of the total box mass experiences origin-crossing by $z=6.8, 6.6, 6$, and $4$, respectively, which matches our simulation results. It is also useful to discuss the time when the cold layer begins to form in the simulation, which we approximate as the 12.5\% mass collapse time ($z_{\rm CLFT}\approx6.6$).

We also run a `late-collapse', $z_{\rm coll}=3$, simulation where $\approx1\%, ~12.5\%, ~25\%$, and $50\%$ of the total box mass experiences origin crossing by $z=5.2, ~5.1, ~4.6$, and $3$, respectively. For the `tophat' profile simulation, collapse is instantaneous, meaning $\approx1\%, ~12.5\%, ~25\%$, and $50\%$ of the total box mass experiences origin crossing at $z_{\rm CLFT}=z_{\rm coll}=6$. The tophat collapse redshift was chosen to match the $25\%$ mass collapse time of the fiducial Gaussian sheet.

\section{Cosmic Sheet Gas Properties and Fragmentation} \label{sec:IGMGasFragmentation}

\begin{figure*}[]
  \centering
  \epsscale{1.0}
  \includegraphics[height=0.35\textheight]{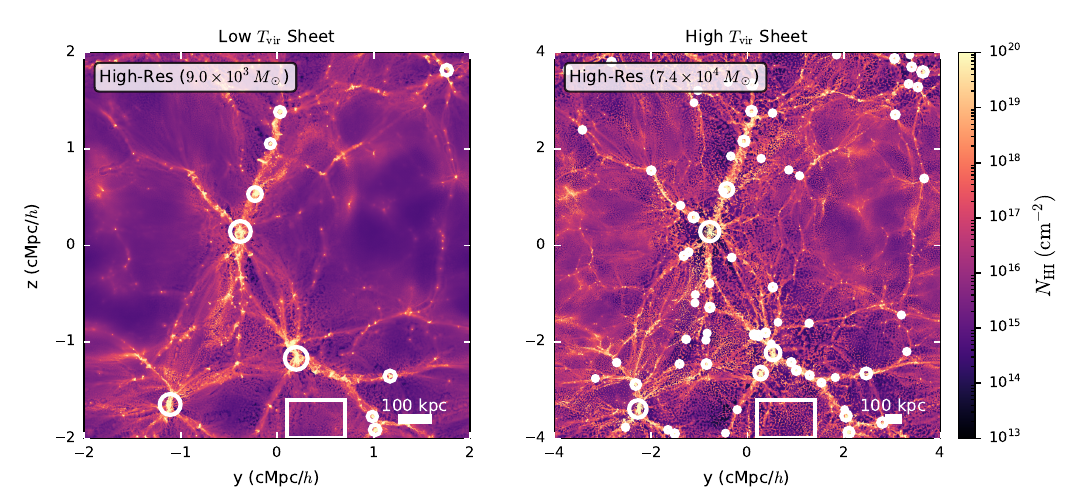}
  \caption{Full-box, face-on projections of neutral hydrogen column density for our high-resolution simulations ($N_{\rm tot}=2\times512^3$). The left and right panels are the low and high \Tvir~sheets, respectively. The baryonic particle mass of the simulation is annotated on each panel. Halos with $M_{\rm h}>10^{10}M_{\odot}$ are marked with circles with radii sized proportionally to $R_{200}$. The white rectangle indicates the IPM subregion shown in Figure~\ref{fig:SubregionProjections}.
  \label{fig:FullSheetProjections}
  }
\end{figure*}

\begin{figure*}[]
  \centering
  \epsscale{1.0}
  \includegraphics[height=0.6\textheight]{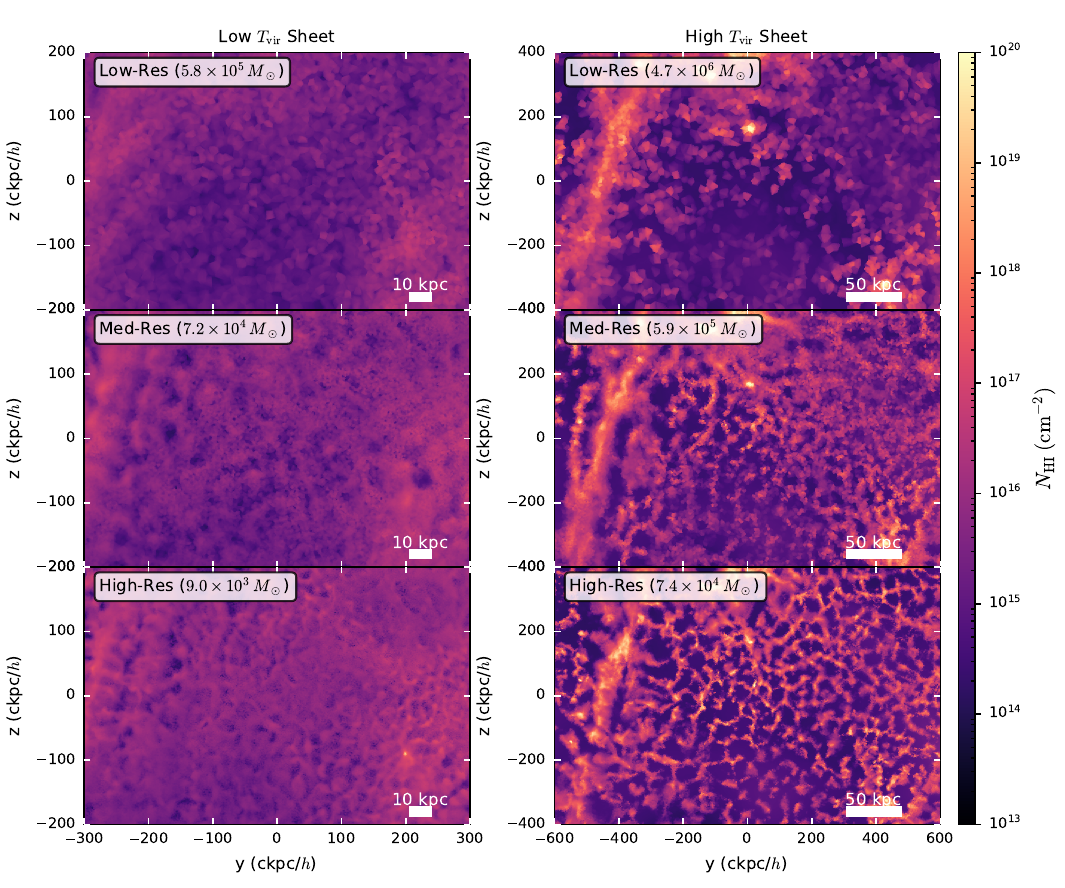}
  \caption{Face-on neutral hydrogen column density projections of the IPM subregion marked in Figure~\ref{fig:FullSheetProjections}, for the low and high \Tvir~sheets (left and right columns). Resolution increases from top to bottom, revealing progressively stronger fragmentation. 
  }
  \label{fig:SubregionProjections}
\end{figure*}

In this section, we explore the kpc-scale morphology of cosmic sheet gas at high resolutions for our fiducial set of simulations. The fiducial set includes the full $\Lambda$CDM matter power spectrum, $P_m(k)=P_{m, \Lambda \rm CDM}(k)$, giving rise to substructure within the sheet. Figure \ref{fig:FullSheetProjections} shows a face-on projection of the high-resolution sheet plane. In the sheet there are three massive halos connected by large cosmic filaments, with many smaller filaments crossing the plane. The substructure of the low (left) and high (right) \Tvir~sheets appears qualitatively similar due to their identical random seeds, but the masses of the filaments and halos of course differ.

Section~\ref{subsec:ResolutionFragmentation} examines the resolution and virial-temperature dependence of sheet fragmentation, confirming \citet{Mandelker2019}.  Section~\ref{subsec:GasMorphology} identifies fragmentation as variations in the thickness and disruption of the cold sheet core. Section~\ref{subsec:N_HI_Enhancement} quantifies the neutral hydrogen column density statistics of fragmenting sheets and their resolution dependence.

\subsection{Resolution Enhanced Fragmentation of the IPM} \label{subsec:ResolutionFragmentation}

Figure \ref{fig:SubregionProjections} shows face-on $N_{\rm HI}$ projections of an IPM subregion (marked in Figure \ref{fig:FullSheetProjections}) for the low and high \Tvir~cosmic sheets (left vs. right columns) across resolutions (top to bottom). The specific subregion was selected for being visibly fragmented and free of halos or filaments; our results are the same across different IPM subregions (see Section \ref{subsec:N_HI_Enhancement}).

The two sheets exhibit distinct neutral hydrogen morphologies. The high \Tvir~sheet (right column) shows strong spatial variations in the neutral hydrogen column density, with localized high-column systems surrounded by low-column regions. The lower virial temperature sheet (left column) has a more spatially uniform and intermediate column density, with small `holes' of lower columns.

The fragmented morphology becomes more pronounced with increasing resolution, especially for the high \Tvir~sheet (right column). Variations in $N_{\rm HI}$ on spatial scales of $ \lesssim 10 ~\kpc$ reveal a rich gaseous morphology largely unresolved in typical cosmological simulations. Despite our efforts to push to higher resolutions, the morphological changes from medium to high resolution demonstrate that neither sheet is numerically converged. \citet{Mandelker2019, Mandelker2021} also reported unconverged IPM fragmentation in a sheet resembling our high \Tvir~case.

\subsection{Kiloparsec-Scale IPM Gas Morphology} \label{subsec:GasMorphology}

\begin{figure*}[p]
    \epsscale{1.0}
    \plotone{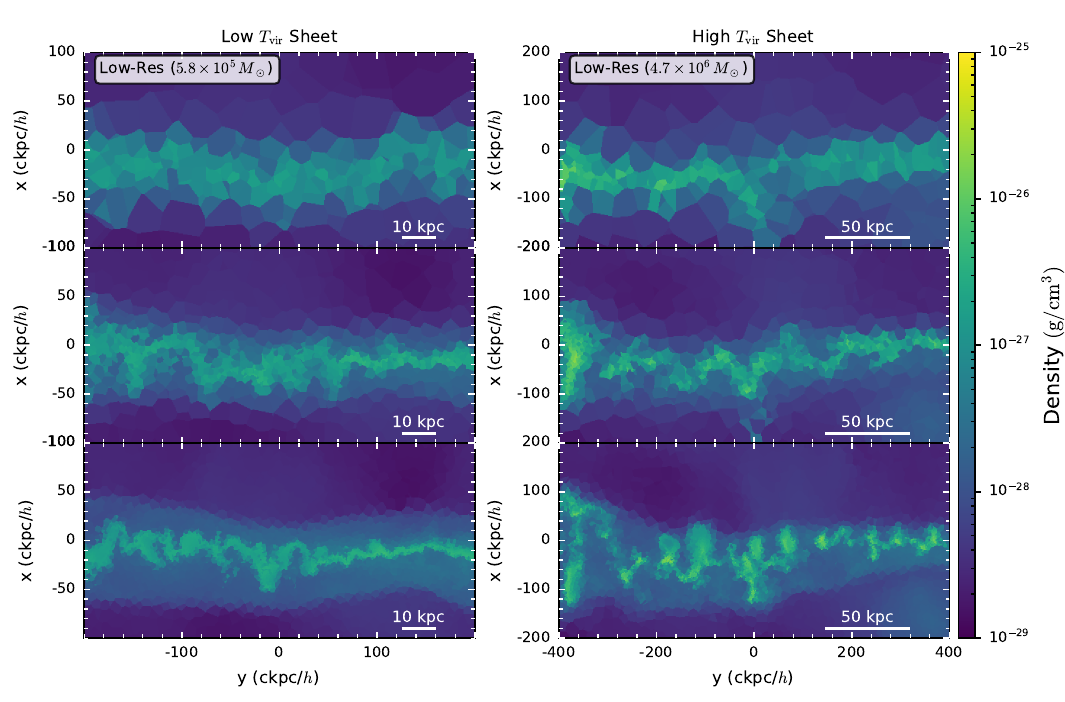}
    \caption{Gas density slice perpendicular to the sheet plane, through the center of the subregion shown in Figure \ref{fig:SubregionProjections}.
    \label{fig:PerpendicularSlice_rho}
    }
\end{figure*}

\begin{figure*}[p]
    \epsscale{1.0}
    \plotone{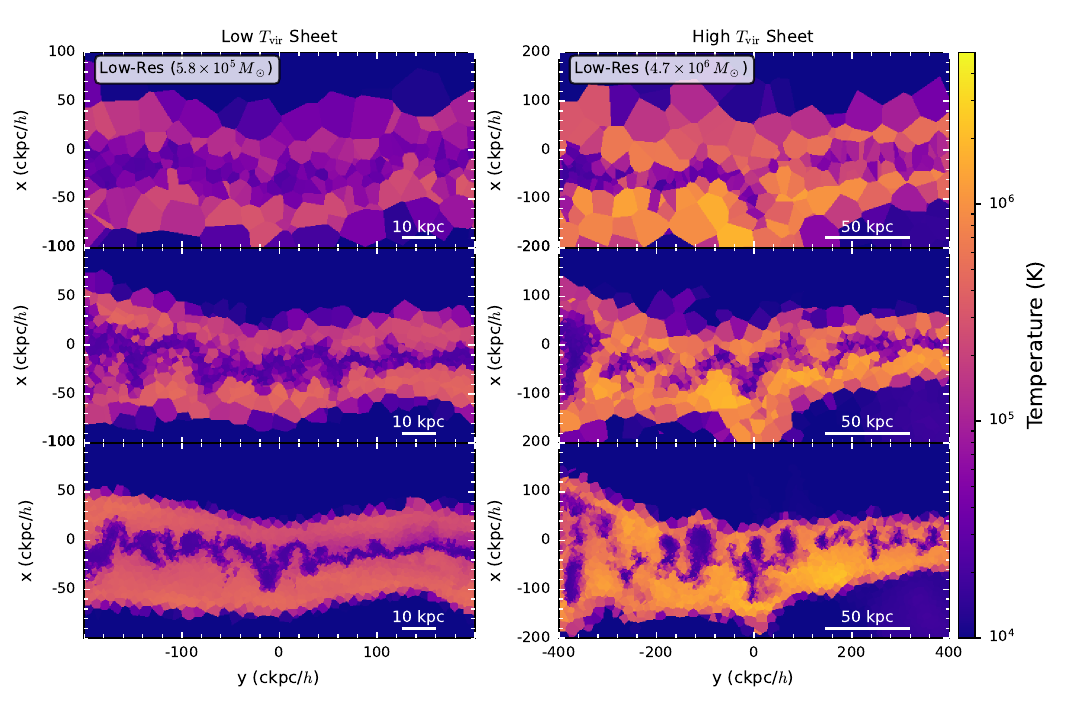}
    \caption{Same as Figure \ref{fig:PerpendicularSlice_rho}, but for gas temperature.
    \label{fig:PerpendicularSlice_temp}}
\end{figure*}

We now examine the gas substructure producing the fragmentation phenomena viewed in projection in the previous subsection. An idealized picture of a cosmic sheet is a central cold core ($10^4~$K) surrounded by hot virialized gas ($T>10^5~\rm K$), bounded by an accretion shock from the supersonic flow onto the sheet. This is the gas behavior in our homogeneous simulation set (not shown). However, the picture is more complex in the fiducial set due to density perturbations. Figures \ref{fig:PerpendicularSlice_rho} and \ref{fig:PerpendicularSlice_temp} show gas density and temperature slices perpendicular to the sheet plane for the low and high $T_{\rm vir}$ sheets.


For our lower $T_{\rm vir}$ sheet (left column), the cold central layer is not well resolved in the low resolution simulation, which has $m_{\rm gas}\approx6\times10^{5}M_{\odot}$ and roughly $\Delta x_{\rm cell} = 3.5-1.6$ pkpc cell-sizes for the $\rho=10^{-27}-10^{-26}~\rm g ~cm^{-3}$ cold gas. Increasing resolution produces a thinner, more corrugated cold layer. At medium resolution, $m_{\rm gas}\approx 7\times 10^{4}M_{\odot}$ ($\Delta x_{\rm cell} = 0.8-1.7 ~\rm pkpc$), we begin to resolve bends and variations in the cold-layer thickness. At high resolution, $m_{\rm gas}\approx 9 \times 10^{3}M_{\odot}$ ($\Delta x_{\rm cell} = 0.4-0.9 ~\rm pkpc$), the cold layer becomes thinner and bent on smaller scales. At all resolutions, the cold layer remains contiguous. The wavelength of these bends in the cold layer decreases with resolution. Combined with better resolving the cold layer thickness, the changing spatial wavelengths of these cold layer bends explain the evolving projected $N_{\rm HI}$ morphology seen in Figure \ref{fig:SubregionProjections}. The projected $N_{\rm HI}$ `holes' are simply locations where the cold layer has become extremely thin and, thus, has little neutral hydrogen along the LOS.

The high \Tvir~sheet (right column) has a more dramatic evolution, with the cold layer becoming disrupted at high resolution. Similar to the low \Tvir~sheet, the cold gas is poorly resolved as a noisy, barely contiguous central layer at low resolution, $m_{\rm gas}\approx 5 \times 10^{6}M_{\odot}$ ($\Delta x_{\rm cell} = 3.2-6.9 ~\rm pkpc$). At medium resolution, $m_{\rm gas}\approx 6 \times 10^{5}M_{\odot}$ ($\Delta x_{\rm cell} = 1.6-3.5 ~\rm pkpc$), the cold layer develops strong bends. At high resolution, $m_{\rm gas}\approx 7 \times 10^{4}M_{\odot}$ ($\Delta x_{\rm cell} = 0.8-1.7 ~\rm pkpc$), the cold layer disrupts into cold clouds. These clouds concentrate in the upper half of the sheet due to the accretion of an extended overdense structure from above at $z\approx4.5$. The clouds produce high-$N_{\rm HI}$ absorber islands in the face-on perspective, as was shown in Figure \ref{fig:SubregionProjections}. Sight-lines through only hot virialized gas produce the low-$N_{\rm HI}$ areas. The resulting $N_{\rm HI}$ distribution is bimodal, which we discuss further in Section \ref{subsec:N_HI_Enhancement}.

A note on terminology: We use `fragmentation' to refer to the sheet having small-scale spatial variations in $N_{\rm HI}$, as is visible in Figure \ref{fig:SubregionProjections}. Fragmentation can manifest from either strong lateral variation in the thickness of the cold central layer, as in the low \Tvir~sheet, or from the full disruption of the cold layer such that cold-clouds are suspended in a hot medium, as in the high \Tvir~sheet. All sheets are multiphase in that there are hot and cold layers, but only the high \Tvir~sheet is strongly multiphase in the sense that the cold gas is in the form of many clouds suspended in a hot medium.

\subsection{Neutral Hydrogen Column Density Enhancement} \label{subsec:N_HI_Enhancement}

\begin{figure}[ht!]
    \epsscale{1.0}
    \plotone{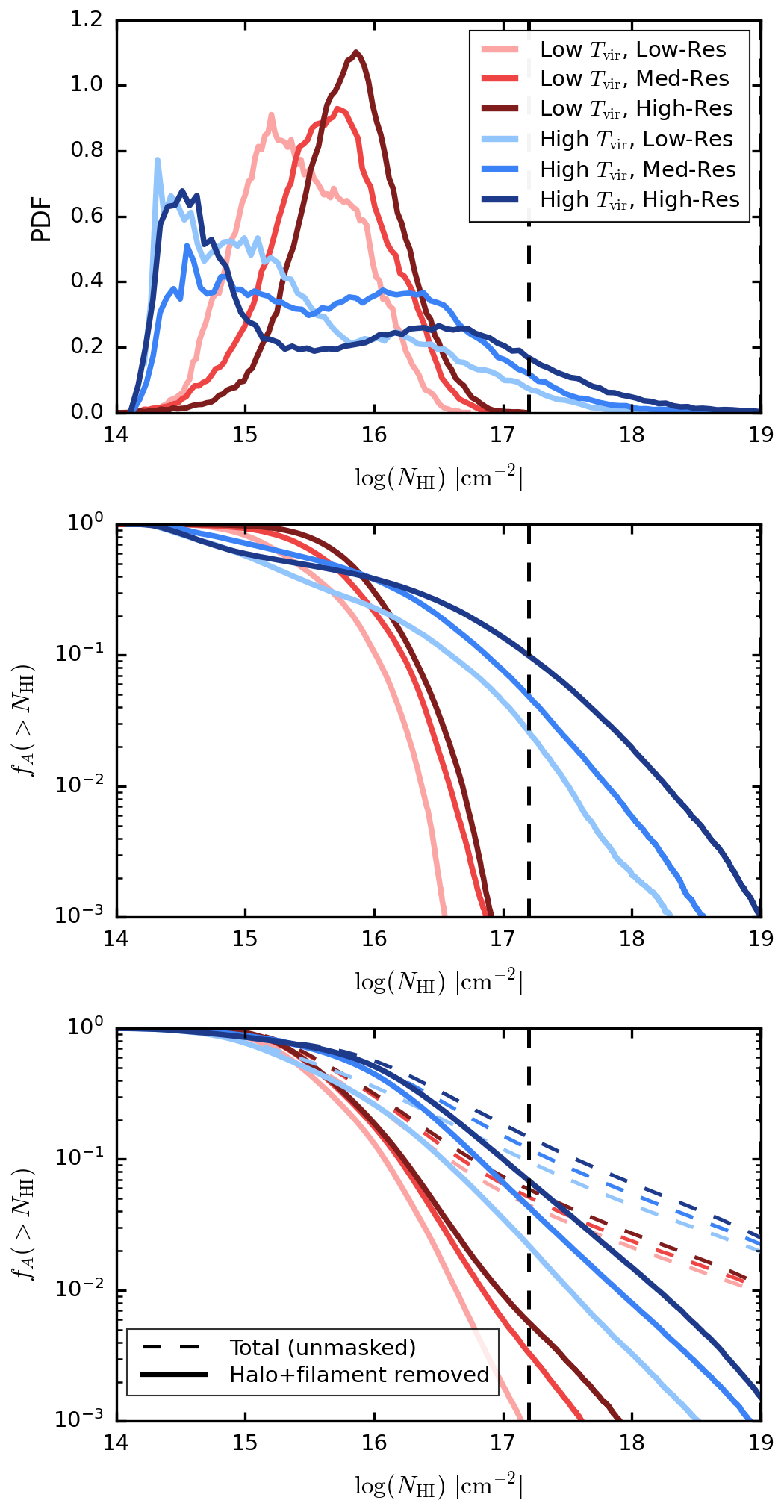}
    \caption{Neutral hydrogen covering fraction distributions in our cosmic sheets. The red and blue colors represent the low and high \Tvir~sheets, respectively. The different shades of each color represent different simulation resolutions, going from light (low res) to dark (high res). \textit{Top panel}: Probability distribution of $N_{\rm HI}$ in the IPM subregion of Figure \ref{fig:SubregionProjections}. \textit{Middle panel}: Cumulative covering fraction distribution (CCFD) of $N_{\rm HI}$ of the IPM subregion. \textit{Bottom panel:} CCFD of $N_{\rm HI}$ across the entire sheet.  The dashed lines are the CCFD across the entire cosmic sheet box, and the solid lines are after masking out the filament and halo regions to capture only the IPM. The vertical gray dashed line denotes the threshold for LLSs at $N_{\rm HI}=10^{17.2}\rm ~cm^{-2}$.
    \label{fig:NHI_ResComp_Subregion}}
\end{figure}

Figure \ref{fig:NHI_ResComp_Subregion} shows the $N_{\rm HI}$ PDF of the IPM subregion (top panel), the $N_{\rm HI}$ cumulative covering fraction distribution (CCFD) of the IPM subregion (middle panel), and the $N_{\rm HI}$ CCFD for the entire sheet (bottom panel). The red and blue lines show the low and high \Tvir~sheets, respectively; the different shades represent different simulation resolutions. Both the high and low \Tvir~sheets show a significant enhancement of pLLS/LLS absorbers with increasing resolution.

The $N_{\rm HI}$ distribution in the IPM region of the low \Tvir~sheet (red lines, top panel) is unimodal, shifting towards higher columns with increased resolution (light to dark). This is caused by an increased mass of neutral hydrogen in the subregion, $M_{\rm HI}=\{1.0, ~1.7, ~2.4\} \times 10^{6}~M_{\odot}$ from low to high resolution at $z=4$.

The additional neutral hydrogen is a result of both a higher average cold-layer density (and therefore neutral hydrogen fraction, $x_{\rm HI}$) and a larger mass of cold-gas overall. Higher resolutions lead to resolving larger gas densities due to increased variance in the density field from small scales and, importantly, less numerical broadening of the central cold layer during collapse. Higher densities increase the cooling rate and $x_{\rm HI}$, enhancing the cold gas and neutral hydrogen mass with higher resolution. 
We find the mechanism generating the majority of our sheets' cold gas is isothermal compression (cold mode accretion) rather than cooling virial layer gas (hot mode accretion). This disfavors the hypothesis that thermal instability is the primary fragmentation mechanism (see Section \ref{subsec:Mechanism}). 

The $N_{\rm HI}$ PDF for the high \Tvir~sheet (blue lines in top panel of Fig. \ref{fig:NHI_ResComp_Subregion}) is bimodal, with peaks near $10^{14.5}~\rm cm^{-2}$ and $10^{16.5}~\rm cm^{-2}$. The behavior with increased resolution is more complex than the low-\Tvir~sheet. As with pLLSs in the low \Tvir~sheet, the LLS covering fraction increases with resolution.

Going from low to high resolution for the high \Tvir~sheet, total neutral hydrogen mass of the region increases as $M_{\rm HI}=\{2.7, ~15.1, ~18.6\}\times 10^8M_{\odot}$ at $z=4$. At low resolution, the cold core and virial layer are very poorly resolved, strongly inhibiting the cooling of sheet gas. Going to medium resolution ($m_{\rm gas}\approx 6\times10^5~M_{\odot}$), there is a large $\approx5\times$ enhancement in HI mass. At high resolution ($m_{\rm gas}\approx 7\times10^4~M_{\odot}$) the cold gas mass increase is smaller, but there is still a strong increase in the LLS columns due to a continued evolution in the cold layer morphology. Unlike the low \Tvir~sheet, the resolution dependence of the $N_{\rm HI}$ PDF for the high \Tvir~sheet is not due exclusively to newly resolved cold gas, but now also due to the increased fragmentation generating high-column clouds.

The CCFD of the IPM subregion gas (middle panel) emphasizes the enhancement of high column density absorbers for cosmic sheets. Going from $\rm low \rightarrow high$ resolution, we find that the IPM subregion in the low \Tvir~sheet has a $10\%\rightarrow30\%$ covering fraction enhancement of pLLS absorbers; and the high \Tvir~sheet has a $3\%\rightarrow10\%$ covering fraction enhancement of LLS absorbers. 

Now looking at the full sheet CCFD (dashed lines in bottom panel of Figure~\ref{fig:NHI_ResComp_Subregion}), the enhancements with resolution are not as large as the isolated IPM subregion. This is expected and consistent with the findings of \citet{Mandelker2019} that filaments and nodes are more converged in their column density distributions than the IPM. 

The majority of dense absorbers will come from within halos and filaments, so we also provide the CCFD for the sheets' IPM regions alone, by masking out filaments and halos (solid lines).\footnote{The masking was performed by cutting sheet areas above $3\times$ the mean DM surface density, chosen by visual inspection to ensure all major and minor halos and filaments were removed.} With halos/filaments removed, the full low \Tvir~sheet has a $13\%\rightarrow19\%$ covering fraction enhancement of pLLS absorbers; the high \Tvir~sheet has a $2\%\rightarrow7\%$ covering fraction enhancement of LLS absorbers. The low \Tvir~sheet appears more converged in the IPM between medium and high resolution than the high \Tvir~sheet, as expected given its higher resolution and lower flow velocities in the low \Tvir~case.

 
To check how representative the IPM subregion shown in Figures \ref{fig:SubregionProjections}-\ref{fig:PerpendicularSlice_temp} is of the full sheet, we can compare the middle and bottom panels of Figure \ref{fig:NHI_ResComp_Subregion}. From low to high resolution, the low \Tvir~sheet pLLS covering fraction enhancement is roughly $3\times$ for the example IPM subregion vs $1.5\times$ for the full sheet. For the high \Tvir~sheet, the LLS enhancement is roughly $3.5\times$ for both. As expected, there is some variation across the sheet as the matter surface density is not perfectly uniform. The subregion is slightly elevated in neutral hydrogen relative to the full sheet. This is somewhat expected as it was chosen visually for its fragmented appearance; we confirmed its $N_{\rm HI}$ enhancement is consistent with random scatter across the sheet and not an outlier. 


\section{Numerical Tests of Sheet Fragmentation}  \label{sec:NumericalTesting}

\begin{figure*}[ht!]
    \epsscale{1.0}
    \plotone{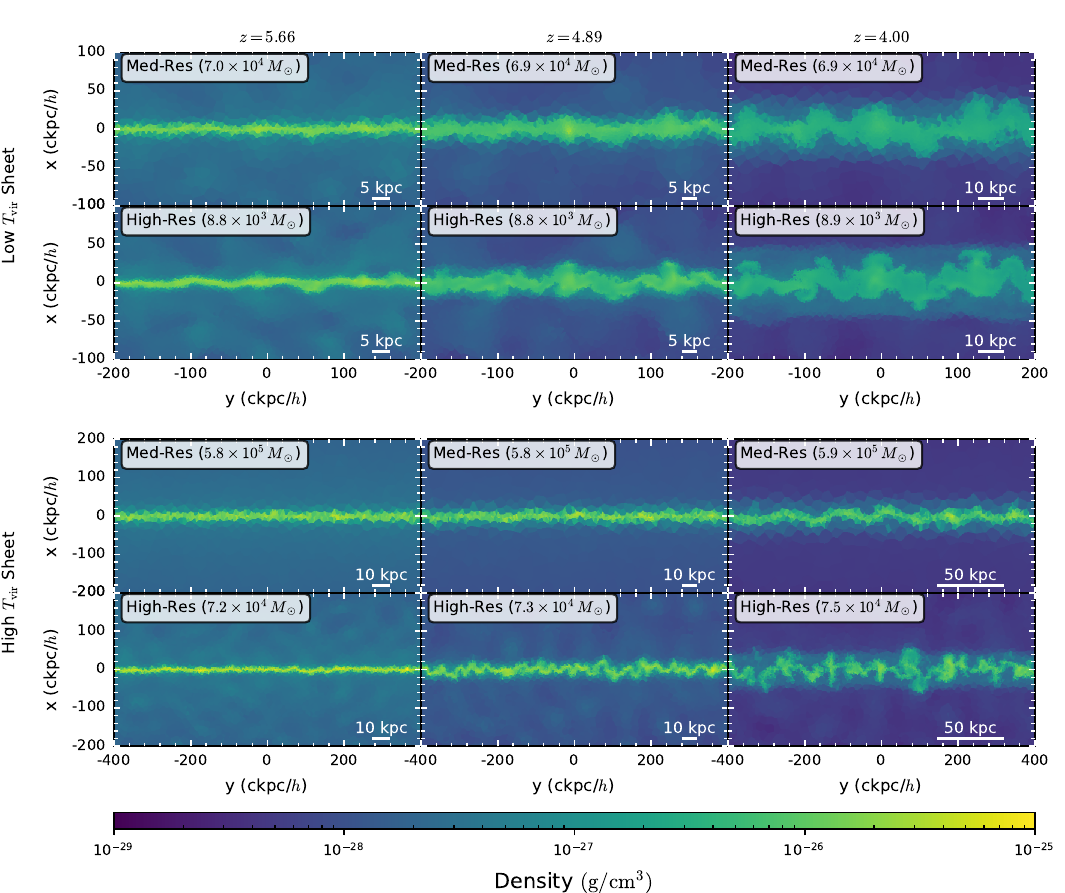}
    \caption{Density slices through a subregion of our high-$k$ perturbed sheets at two different resolutions. The three columns show different redshifts, $z\approx\{5.66, 4.89, 4.00\}$. The first and second rows are the medium and high resolution simulations of the low \Tvir~sheet. The third and fourth rows are the medium and high resolution simulations of the high \Tvir~sheet.
    \label{fig:Highk_RhoSlice}}
\end{figure*}

\begin{figure*}[ht!]
    \epsscale{1.0}
    \plotone{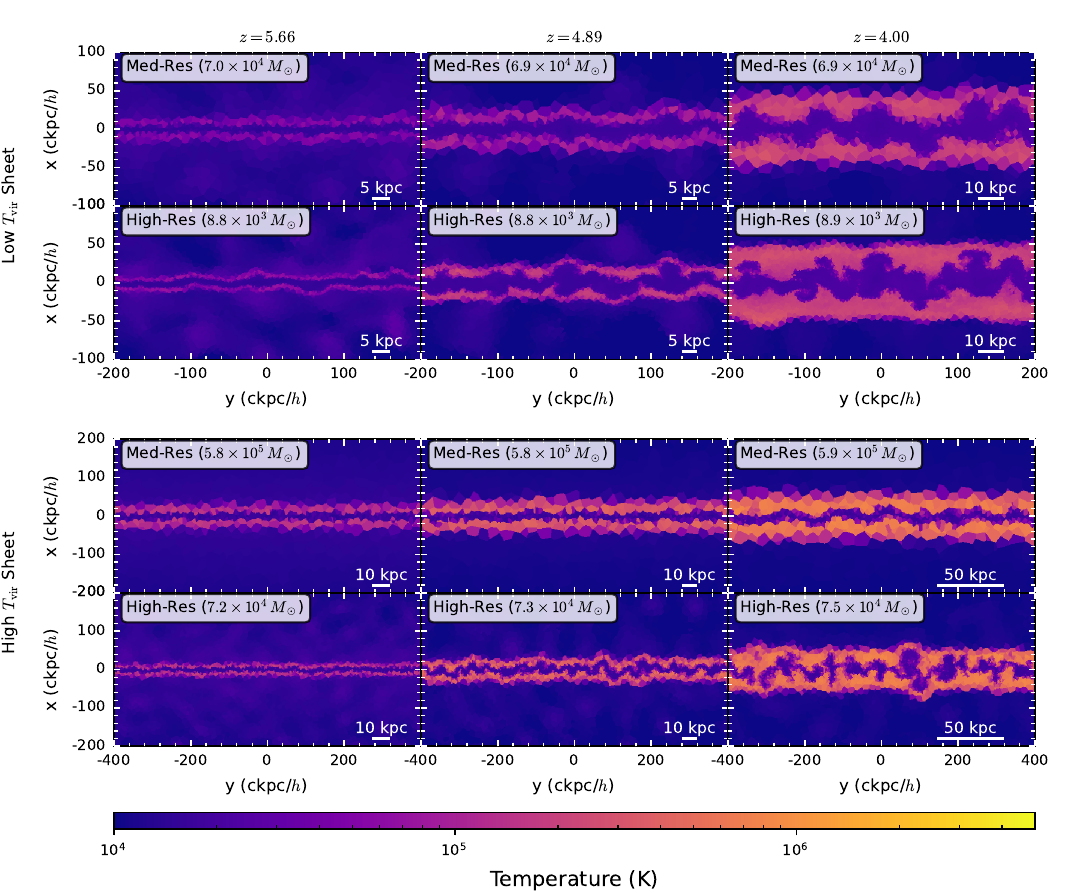}
    \caption{Same as Figure \ref{fig:Highk_RhoSlice}, but for the gas temperature.
    \label{fig:Highk_TempSlice}}
\end{figure*}

We investigate the physical mechanism behind sheet fragmentation using our high-$k$ and homogeneous simulation sets. We find the homogeneous cosmic sheets do not fragment, suggesting that initial perturbations are required to seed fragmentation. Thus, in this section we focus on the results from the high-$k$ perturbation set where we do observe fragmentation.

The simulations reveal that fragmentation arises due to the growth of bending-mode perturbations in the sheet's cold layer, seeded by the inhomogeneity of the accretion flow (Section \ref{subsec:UnstableEvolution}). The gas accretion-flow vectors demonstrate that these bending mode perturbations grow via the lateral deflection of the accreting gas into the peaks and troughs of the seeded-bends (Figure \ref{fig:StreamSlices}). We examine how variations in the cosmic sheet properties, such as late collapse and an alternative density profile, affect fragmentation (Section \ref{subsec:ModifiedCollapse}). Across our simulations, the growth of these bending modes is consistent with nonlinear thin shell instability (NTSI), also known as the Vishniac instability \citep{Vishniac1994}. We focus here on the qualitative picture for fragmentation via NTSI using the simulations, and later provide the analytic details of growth timescales and wavelengths in Section \ref{subsec:NTSItheory}.

\subsection{Fragmentation by Cold-Layer Bending Modes} \label{subsec:UnstableEvolution}

Fragmentation manifests in our high-$k$ simulation set as a buckling of the sheet's central cold layer. Figures \ref{fig:Highk_RhoSlice} and \ref{fig:Highk_TempSlice} show the density and temperature evolution ($5.66>z>4$) of our high-$k$ perturbed sheets at two resolutions for both sheets. The formation and evolution of the virial layer and cold layer are visible in these figures.

The cold layer initially forms at $z\approx6.6$, becoming shock bounded around $z=6.14$. The cold layer at $z=5.66$ (left column) is shock bounded, with the shocks visible as single-cell-thick layers of warm gas. Efficient post-shock cooling initially prevents the growth of a thick virial layer. After $z=4.89$ (middle column), the virial layer starts to grow as the post-shock gas cooling time becomes large relative to the accretion time. By $z=4$ (right column) the virial layer is now thick. Meanwhile, bending perturbations are also evolving. Small bending modes are present at $z=5.66$, seeded by the inhomogeneous accretion flow. By $z=4.89$ we observe these bending perturbations have grown in amplitude, which we will demonstrate is consistant with NTSI. At late times, $z=4$, the cold layer gains more complex morphologies, which we will show is due to shear and vorticity driven by gas flows in the newly formed virial layer.

There are strong resolution-dependent variations in the time-evolution of the sheets. Visually, one can see numerical noise in the medium resolution simulation cases. This demonstrates sheets are very poorly resolved in big-box simulations. For the high \Tvir~sheet, the bending perturbation growth between $z=6.14-4.89$ is largely suppressed in the medium resolution case. Additionally, the gas flows within the newly formed virial layer become important at high resolution. This manifests in the low \Tvir~sheet as wave-like features in the cold layer, and more aggressively in the high \Tvir~sheet as the disruption of the cold layer. 

\begin{figure*}[ht!]
    \epsscale{1.0}
    \plotone{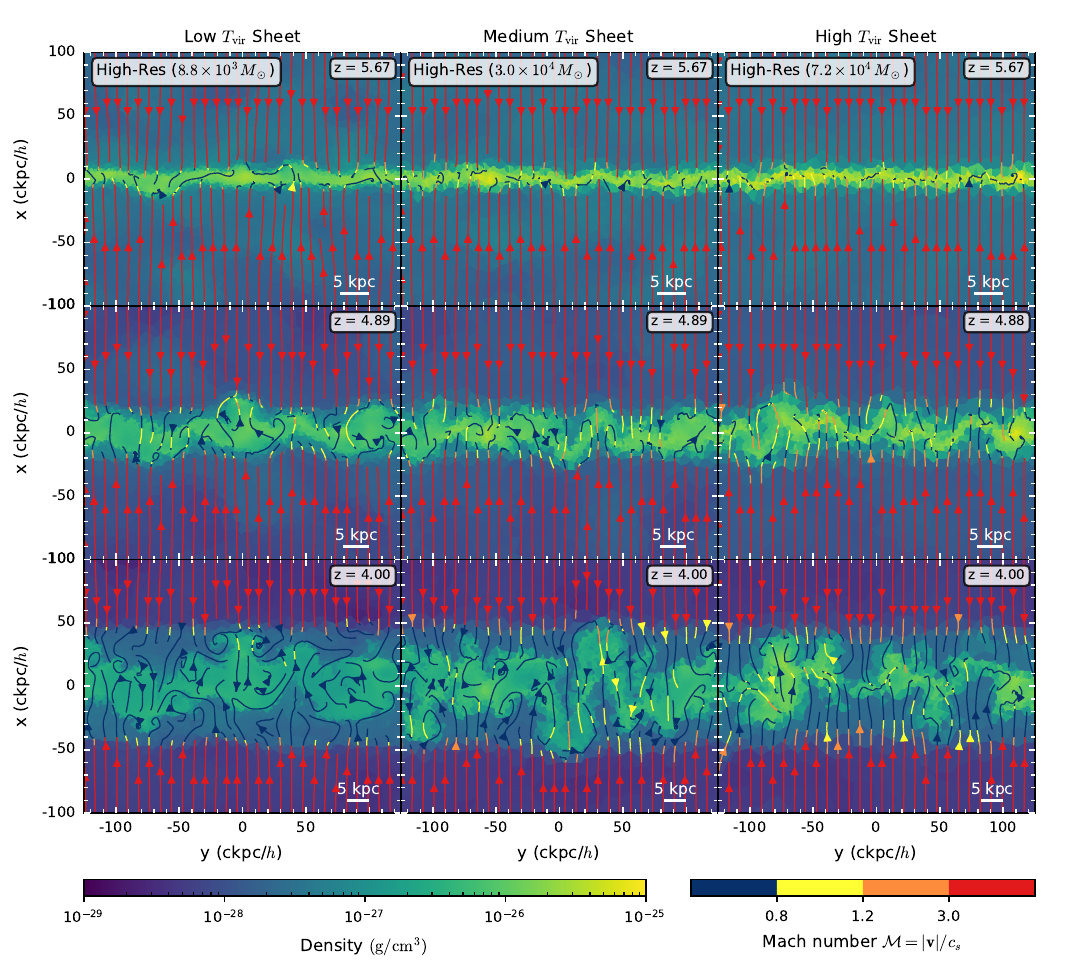}
    \caption{Gas streamlines overlaid on density slices for the low-, medium-, and high-\Tvir~sheets (left to right columns) at three redshifts ($z=5.67, ~4.89,~ 4.00$). The medium \Tvir~sheet corresponds to a box size of $L=6~\cMpc/h$ with \Tvir$\approx4\times10^5$ K. The streamline colors represent the Mach number of the flow: blue = subsonic, yellow = transonic, orange and red=supersonic.
    \label{fig:StreamSlices}}
\end{figure*}

In Figure \ref{fig:StreamSlices} we show the gas dynamics of the inflow across time for our high-resolution simulations using stream plots. The color of the inflow lines denotes the Mach number, from subsonic (blue) to transonic (yellow) to supersonic flow (red/orange). We now summarize the qualitative picture of cosmic sheet fragmentation through the three time-stages in Figure \ref{fig:StreamSlices}:

The first stage of cosmic sheet fragmentation is shown in the top row of Figure \ref{fig:StreamSlices} ($z=5.67$). The cold-overdense layer forms via cold-accretion, and is then bounded on either side by an accretion shock. The inhomogeneous nature of the accreting gas, due to the $\Lambda$CDM matter power spectrum, seeds small bending-mode perturbations in the central cold layer. The accreting gas shocks onto the cold layer at slightly oblique angles due to the local bends in the sheet surface. This is a traditional setup for NTSI. 

The oblique impact of the inflowing gas transfers lateral momentum, concentrating inflow into the peaks and troughs of the bends in the cold layer. The net momentum flux of gas into the peaks and troughs grows the bending perturbation amplitude. By $z=4.89$ (middle row), NTSI has provided some growth to the originally small bending perturbations. An increase in the amplitude of the bending perturbations increases the obliquity of the accretion angle, which subsequently increases the growth rate of the cold-layer bending amplitude until saturation. 

By $z=4$ (bottom row, Fig. \ref{fig:StreamSlices}), traditional NTSI is saturated or cut off by the formation of a virial layer. In the transition from shock bounded to virial-layer bounded, the cold layer experiences shearing flows from the growing hot gas layer. This leads to vorticity and Kelvin-Helmholtz instability (KHI), as is common for late-stage NTSI, which further disrupt the sheet \citep{Blondin1996, Heitsch2007}. For the low \Tvir~sheet (bottom-left panel), inflowing gas is still being directed into troughs, but now subsonically, and the shear of the inflow off the cold layer generates vorticity. For the high \Tvir~sheet (right column), NTSI saturates causing the cold layer to have bends with amplitude of roughly the entire sheet thickness (i.e., to the edge of the virial shock). Fast post-shock gas gets deflected and concentrated by the asymmetric morphology of the cold layer. These high velocity flows contribute to a turbulent medium which further disrupts the cold gas.

\subsection{Other Modified Sheet-Collapse Experiments} \label{subsec:ModifiedCollapse}

\begin{figure*}[ht!]
    \epsscale{1.1}
    \plotone{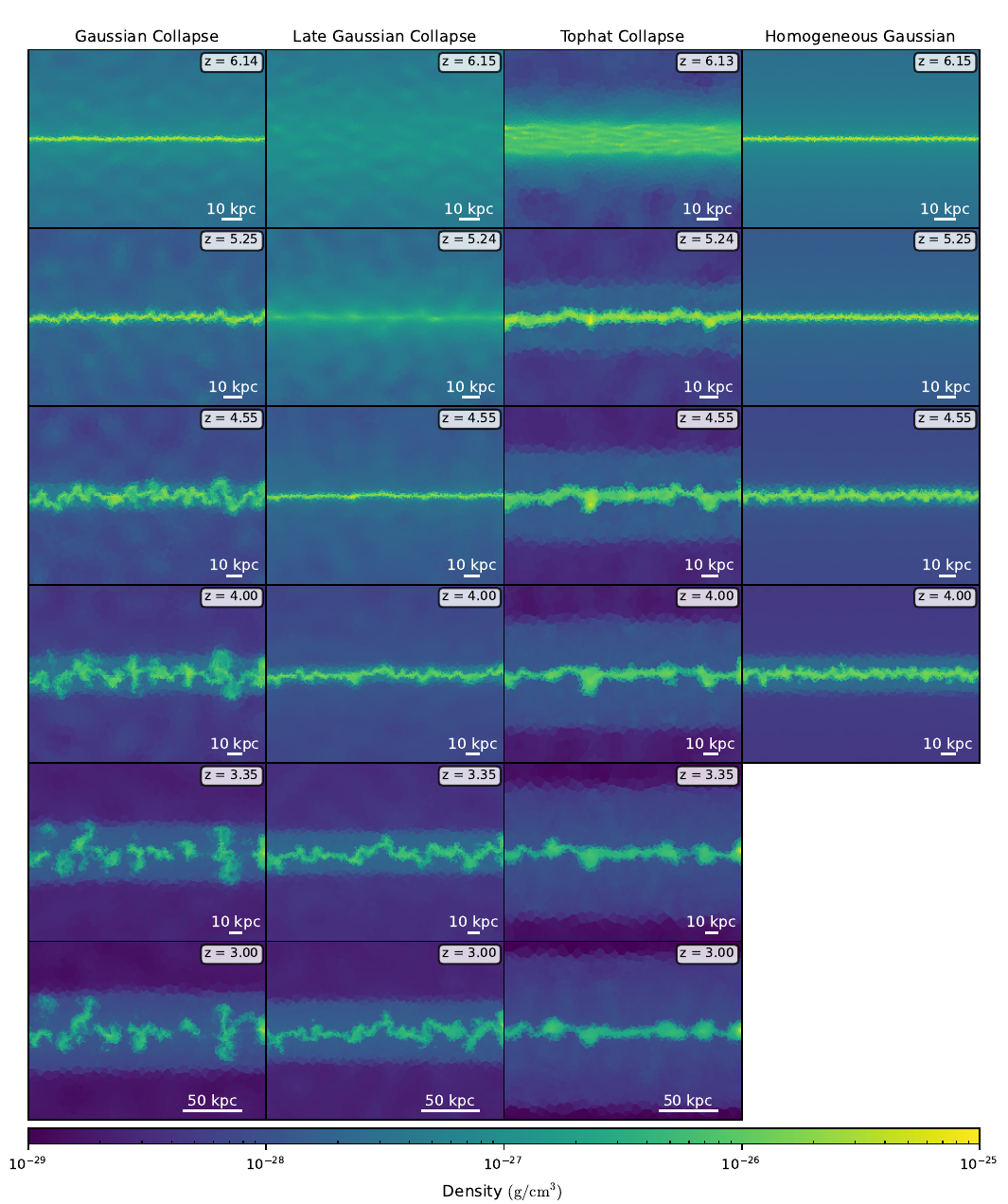}
    \caption{Evolution of gas density slices in four high-resolution, high-\Tvir~sheet simulations of varying collapse histories. Columns show, from left to right: the fiducial Gaussian sheet with high-$k$ perturbations; delayed collapse Gaussian ($z_{\rm coll}=3$, $z_{\rm CLFT}\approx5.1$); a tophat profile ($z_{\rm coll}=6$); and the fiducial Gaussian sheet without background perturbations from the homogeneous set. All panels span $400\times300~\ckpc/h$.
    \label{fig:AlternativeScenarios}}
\end{figure*}

We now offer insight into how our results generalize to a broader parameter space of cosmic sheets using the following high-resolution simulation scenarios:
\begin{enumerate}[(a)]
    \item A medium \Tvir~sheet (Fig. \ref{fig:StreamSlices}, middle-column). 
    \item A delayed-collapse high \Tvir~sheet, with $z_{\rm coll}=3, ~z_{\rm CLFT}\approx5.1$ (Fig. \ref{fig:AlternativeScenarios}, 2nd-column). 
    \item A high \Tvir~sheet with a tophat density profile (Fig. \ref{fig:AlternativeScenarios}, 3rd-column).
    \item The homogeneous set version of the fiducial high \Tvir Gaussian sheet (Fig. \ref{fig:AlternativeScenarios}, 4th-column).
    \item Late time evolution to $z_{\rm end}=3$ of these fiducial and modified sheet simulations (Fig. \ref{fig:AlternativeScenarios}, bottom row). 
\end{enumerate}

In the middle column of Figure \ref{fig:StreamSlices}, we show a medium virial temperature, $T_{\rm vir}=4\times10^5~\rm K$, sheet. It uses a box of size $L=6~\rm cMpc/h$. The sheet's behavior is qualitatively quite intermediate between the low and high \Tvir~sheets, which is expected. 

While previous figures focused on $z\geq 4$, the bottom two rows of Figure \ref{fig:AlternativeScenarios} show the evolution down to $z=3$. NTSI growth is suppressed at this redshift because the virial layer has matured; shearing flows weaken as the virial layer thickens and thermalizes. Evolution is instead driven by cosmic expansion and passive cooling. The cold clouds in the IPM appear long-lived, in the sense that they persist from $z=4\rightarrow3$. The true long-term survival, in the turbulent IPM and/or against evaporation, remains unknown \citep{Gronke2022}. Whether these large cold clouds survive accretion into filaments or halos and the subsequent effect on filament/halo cold gas supplies is an area for future work.

The 2nd column of Figure \ref{fig:AlternativeScenarios} shows the late-collapse Gaussian sheet scenario ($z_{\rm coll}=3$, $z_{\rm CLFT}\approx5.1$). Clearly visible is the delayed formation of the cold layer. By $z=3$, the late-collapse scenario has experienced growth in the amplitude of the bends, but there has not been disruption of the cold layer into many individual clouds. The late collapse case is more weakly fragmented, even comparing the sheets at times of equivalent sheet mass.

The 3rd modified sheet scenario is the tophat initial density profile. Neither our fiducial Gaussian nor this tophat profile perfectly represent all cosmic sheets. Cosmic sheets collapse from locally anisotropic regions of the primordial density field, where the superposition of Fourier modes from the matter power spectrum produces coherent compression along a single principal direction. Consequently, such structures are not well described by any single idealized overdensity profile (e.g., Gaussian or tophat). Thus we present here the (somewhat extreme) tophat scenario to demonstrate how fragmentation varies when changing our fiducial Gaussian profile assumption. 

The tophat profile sheet collapses nearly instantaneously at $z_{\rm coll}=z_{\rm CLFT}=6$ (Fig. \ref{fig:AlternativeScenarios}, 3rd column). This results in a more massive cold layer compared to the Gaussian cases because the 50\% box-mass collapse occurs at a time of higher mean cosmic density. The thicker cold layer in the tophat case has a longer NTSI growth time than the Gaussian case due to the larger inertia of the cold layer (see Eqn. \ref{eqn:NTSI_GrowthTime}). Additionally, the virial layer forms nearly instantaneously after the cold layer forms, meaning there is no NTSI growth. The resulting cold layer still has lateral thickness variations, which are not breathing-modes but rather spherical overdensities caused by small DM halos formed prior to collapse from the $\approx100~\ckpc/h$ perturbation modes.\footnote{This also occurs in the Gaussian case but is exacerbated by the tophat sheet reaching higher densities at earlier times.} While the Gaussian initial profile is a more realistic cosmic sheet growth model than the tophat, this demonstrates that cold layer morphology depends on profile choice and we thus recommend future work exploring a range of sheet profiles taken from large-volume cosmological simulations.

In the 4th column of Figure \ref{fig:AlternativeScenarios} we show the case of a homogeneous Gaussian sheet, i.e. $P(k)=0$ except for the 1D-Gaussian perturbation. This simulation was only run down to $z=4$. We show this simulation because of the slight perturbations observed at $z=4$, which is unexpected for a perfectly symmetrical and unperturbed inflow. The cause could be numerical noise or carbuncle instability, which we infer as being capable of generating some level of seed perturbations in the cold layer. NTSI will grow even for a homogeneous inflow if seed perturbations are present. Future work at higher resolutions will be needed to definitively confirm the contributions of physical vs. numerical mechanisms in bending perturbation growth for high \Tvir~sheets. We observe no such behavior in the low \Tvir~homogeneous sheet.

\section{Discussion} \label{sec:Discussion}

Cosmic sheets are poorly resolved at large-box simulation resolutions, $m_{\rm gas} \gtrsim 10^5 ~M_{\odot}$, as evidenced by resolution-dependent gas structure and HI absorption statistics (Section \ref{sec:IGMGasFragmentation}, Figs. \ref{fig:PerpendicularSlice_rho}-\ref{fig:NHI_ResComp_Subregion}). Our analysis revealed that bending instabilities drive fragmentation of the central cold layer (Section \ref{sec:NumericalTesting}, Figs. \ref{fig:Highk_RhoSlice}-\ref{fig:StreamSlices}). We now discuss the evidence for and against possible fragmentation mechanisms (Section \ref{subsec:Mechanism}). We then discuss the prospects for observing the enhanced neutral hydrogen absorption predicted for high-$z$ cosmic sheet cold layers (Section \ref{subsec:ObservableLLSs}). Finally, we briefly discuss our SPH simulations (Section \ref{subsec:Arepo_v_MPgadget}). 
 
\subsection{Mechanism of Sheet Fragmentation} \label{subsec:Mechanism}

We have proposed that nonlinear thin shell instability (NTSI) drives the bending mode growth that fragments the cosmic sheet. Here we recap analytic NTSI theory (Sec. \ref{subsec:NTSItheory}), showing that cosmic sheets satisfy its criteria and that its predicted growth times and wavelengths are consistent with our simulated sheets. We then explain why we disfavor the previously proposed thermal instability (TI) mechanism and Rayleigh-Taylor (RTI) and Richtmyer-Meshkov (RMI) instabilities (Section \ref{subsec:AlternateMechanisms}).

\subsubsection{NTSI Growth Analysis} \label{subsec:NTSItheory}

We apply the analytic results of \citet{Vishniac1994} to estimate whether the NTSI growth timescales and growth wavelengths are consistent with the perturbation growth in our cosmic sheets. The traditional NTSI theory, as derived in \citet{Vishniac1994}, describes a cold-slab of gas bounded on either side by a shock from an incident isothermal supersonic flow. The bend-amplitude of the cold-slab, $\xi_{\rm amp}$, is defined as the displacement between the global sheet midplane and any local deflection peak (or trough) of the cold-slab. The cold-slab enters the `non-linear' bending regime when the bend-amplitude is comparable to the cold-slab thickness, $h$. Given $\xi_{\rm amp}\geq h$, the bend-amplitude of a given $k$-mode has the growth time:

\begin{equation}
    t_{\rm NTSI} (k)\approx \frac{1}{c_sk^{3/2}\xi_{\rm amp}^{1/2}} ~~~,
    \label{eqn:NTSI_GrowthTime}
\end{equation}
where $c_s$ is the sound-speed in the cold-gas slab and $k$ is the perturbation angular wavenumber. Equation \ref{eqn:NTSI_GrowthTime} shows that larger $k$-modes (i.e., smaller wavelengths) grow faster. The growth time was originally derived only as a proportionality, but simulations motivate a proportionality constant near one, which we adopt here \citep{Heitsch2007}.

The growth is faster for larger $k$-modes, but the small-angle approximation used to derive the expression breaks down around $k\xi_{\rm amp}\gtrsim1$ \citep{Vishniac1994}. Plugging in a seed perturbation $\xi_{\rm amp,0}=h$ and the corresponding small-large angle transition k-mode, $k_{\rm max}=1/h$, into Eqn \ref{eqn:NTSI_GrowthTime}, we can find a characteristic NTSI timescale,

\begin{equation}
    t_{\rm NTSI, min}(k_{\rm max})\approx 
    \frac{h}{c_s}~~~.
    \label{eqn:NTSI_MinGrowthTime}
\end{equation}
This timescale is for the fastest growing $k$-mode at the moment it crosses from the small to large angle limit. All larger $k$-modes are in the large-angle bending limit and thus not described well by the analytic theory. All smaller $k$-modes have a longer growth time (proportional to $k^{-3/2}$).\footnote{Not all $k<k_{\rm max}$ grow; there is a lowest growing $k$-mode set by accretion damping \citep{Vishniac1994}. Modes suppressed by accretion damping are at longer wavelengths than those discussed here.} We note that $k_{\rm max}$ is not a precise threshold. Growth at higher bending angles is possible; \citet{McLeod2013} found the saturation scale was closer to $kh\approx\pi$.

We can compare the NTSI growth time to the duration over which the sheet is theoretically unstable to NTSI as a check of our hypothesis. We extract the sheet 1D gas density, temperature, pressure, and velocity profiles from our $P_m(k)=0$ homogeneous cosmic sheet simulation set. These are unperturbed references where it is easy to measure $h$ and check the following NTSI conditions. NTSI will start growing when the cold slab is fully formed and is shock bounded, around $z\approx6.1$. NTSI stops growing when the post-shock layer begins to thermalize because the post-shock gas temperature is large enough that the cooling time becomes longer than the accretion time. The virial layer gains finite thickness around $z\approx4.9$. This gives around 300 Myr for NTSI growth. We note that the gas properties within the virial layer are not well-converged, leaving some ambiguity as to the precise cutoff time. 

For our sheets, $h$ is between $1-4$ kpc for the redshift range where the slab is shock bounded. The cold layer of the high \Tvir~sheet is initially thinner than the low \Tvir~sheet. This range of cold-core thicknesses gives an NTSI growth wavelength around $\lambda_{\rm min}=2\pi/k_{\rm max}=2\pi h=6\text{–}25\,\mathrm{kpc}$, comparable the simulation results. The sound speed is roughly constant with $c_s \approx \rm 20~ km ~s^{-1}$. The resulting theoretical NTSI growth timescale for our fiducial Gaussian sheets is around $50-200~\rm Myr$, increasing with time. With an NTSI growth duration longer than the NTSI growth timescale, we find our cosmic sheets are plausibly undergoing NTSI bending mode instability.

The shorter growth time in the high \Tvir\ sheet is caused by its thinner cold layer at early times, contributing to the stronger perturbation growth observed for our high \Tvir~sheet over the low \Tvir~sheet. The thickness of the cold layer is at the cell-size limit for the high \Tvir~sheet at the time during and just after collapse ($z>6$), even in our highest resolution simulation. Higher resolutions better resolve the narrow density peak during collapse, yielding a stage with a thinner cold layer. This also explains the resolution dependence of $N_{\rm HI}$, as the cooling rate is roughly proportional to density squared.

Even though NTSI growth shuts off at the formation of the virial layer, we still observe some perturbation growth in our simulations at later times. NTSI is driven by accreting gas cooling rapidly, such that inflowing gas cools into the cold slab where deflection by the oblique bending angle yields lateral momentum transport into the peaks and troughs. We observe a qualitatively similar effect for the hot gas flows in the virial layer. The ram pressure component of the post-shock hot gas is deflected into the peaks and troughs, asymmetrically pushing on the cold slab. This is especially prevalent at high \Tvir. The hot gas flows likely need additional numerical resolution to be fully resolved.

The standard NTSI growth-rate estimate assumes an isothermal, non-self-gravitating slab in the small-bending-angle regime \citep{Vishniac1994}. Self-gravity generally has only a modest effect on hydrodynamic slab fragmentation initially, although it can increase the central slab density thereby slightly shortening NTSI growth times \citep{Hunter1986,Hueckstaedt2003}. NTSI requires strong compression because high thermal pressure stiffens the slab against bending; consequently, the adiabatic limit can suppress NTSI, whereas isothermal and radiative cooling flows produce broadly similar morphologies \citep{Klein1998}. Compared to isothermal flows, radiatively cooling colliding flows can exhibit more violent early evolution, while hot post-shock gas exerts a damping effect at later times \citep{Folini2009}. Insufficient cooling can inhibit NTSI, while excessively rapid cooling can excite small-scale instabilities that disrupt coherent bending modes \citep{Hueckstaedt2003}. The strong-to-weak cooling regime transition is difficult to quantify. Hence, we adopt the formation of a thermally dominated virial layer as our practical NTSI cutoff. NTSI-generated shear can become Kelvin-Helmholtz unstable as corrugations steepen \citep{Vishniac1994, Blondin1996, Heitsch2007}, consistent with the late-time KHI in our cosmic sheets (bottom row of Figure~\ref{fig:StreamSlices}).

Lastly, we note that \citet{Anninos1995} ran early 2D simulations of cosmic sheets using grid hydrodynamics, reaching kpc-scale resolutions with log-binned cell sizes. They found NTSI as a dominant instability when relaxing the plane-symmetry constraint in their Section 4. Our results look qualitatively similar.

\subsubsection{Disfavored Instabilities} \label{subsec:AlternateMechanisms}

Fragmentation in our cosmic sheets is a result of bending mode perturbations in the cold layer, which we argued is due to NTSI. \citet{Mandelker2019, Mandelker2021} first observed cosmic sheet fragmentation in a cosmological zoom, with a virial temperature near our high-\Tvir~case. However, their proposed mechanism was thermal instability (TI) \citep{Field1965, Balbus1986}. There is extensive literature on the TI of diffuse astrophysical gas at $T\approx10^{4.5}-10^7~\rm K$, and its ability (or inability) to produce cold clouds \citep[e.g.][]{Field1965, McCourt2012, Waters2019, Das2021, Waters2023}. Under some conditions, thermally unstable gas clouds can fragment via RMI and RTI as a result of rapid collapse after thermally unstable cooling \citep{Gronke2020, Yao2025}.

However, we argue that TI, and subsequently RMI and RTI, are not the sources of fragmentation for cosmic sheets in our simulations, for several reasons: 
(1) We observe instability in the sheet before a virial layer forms. 
(2) The majority of cold gas in our simulations is generated by isothermal compression during initial collapse (cold-accretion), not by cooling down from the virial temperature. We examined the origin of gas ending up in cool clouds, and found that, by $z=4$, only around 25\% of cold gas had cooled from a hot state. 
(3) At late times, the virial layer spans only $\mathcal{O}(10)$ cells even at our highest resolution, which may suppress the RMI/RTI-driven fragmentation studied at higher resolutions in \citet{Gronke2020} and \citet{Yao2025}. 
(4) We find that the difference in HI gas mass ($\Delta M_{\rm HI}$) between medium and high resolution simulations does not increase after the formation of the virial layer (not shown). If a cooling bottleneck opened up due to resolving TI, $\Delta M_{\rm HI}$ between resolutions should increase after the virial layer forms. 
(5) We ran a simulation without helium (not shown), removing the isochorically stable and isobarically unstable region where \citet{Das2021} identified the potential for fragmentation. We observed qualitatively identical behavior to our other sheets, though this was using SPH, with limitations we discuss in Appendix \ref{appendix:SPH}. 
(6) The lack of cold-layer evolution from $z=4\rightarrow3$ is not consistent with TI, but is consistent with NTSI. In TI, more cold gas production is expected once the virial layer is fully developed. 

While our simulations favor NTSI over these mechanisms on resolved scales, their possible contributions at smaller scales remain to be explored in future higher-resolution studies. 

\subsection{The Cosmic Prevalence of pLLS \& LLS with Cosmic Sheet Origin} \label{subsec:ObservableLLSs}

Poor spatial resolution of cosmic sheets in typical large-box simulations leaves their contribution to HI absorber statistics, a key IGM observable, uncertain. Sheets have a much lower mean density than filaments or halos, but have a much larger volume-filling fraction. We now make an order-of-magnitude estimate for the fraction of observed LLS and pLLS absorbers that originate from cosmic sheets at $z=4$. We use our simulations as a high resolution sheet reference and TNG100 to characterize the cosmic sheet population \citep{Nelson2019}. 

We estimate the LOS number density of HI absorbers originating from cosmic sheets, $\ell_{\rm abs}^{\rm sheet}$, as
\begin{equation}
    \ell_{\rm abs}^{\rm sheet}(>N_{\rm HI}) =
    \frac{1}{2L^3}
    \sum_i A_{i}~
    f_{\rm A}(>N_{\rm HI}\mid T_i),
\label{eqn:sheet_abs_incidence}
\end{equation}
where the 1/2 factor accounts for angular averaging, $L^3$ is the TNG100 box volume, the sum is over all the individual cosmic sheets of surface area and shock temperature (A, $T_{\rm shock}$), and $f_{\rm A}$ is the CCFD of HI absorbers for an individual cosmic sheet as a function of virial temperature. We discuss how we define $f_{\rm A}$ and identify all the cosmic sheets within TNG100 below.

The parameter $f_{\rm A}$ is extracted from our high resolution simulations (Fig. \ref{fig:NHI_ResComp_Subregion}, bottom panel, solid lines). The \Tvir~dependence is taken to be a simple power-law fit between our low and high \Tvir\ simulations:
\begin{align}
    f_A(>N_{\rm HI} \mid T_{\rm vir}) &= f_1 
    \left(\frac{T_{\rm vir}}{2\times10^5~\rm [K]}\right)^\alpha; &\alpha &= \frac{\ln(f_2/f_1)}{\ln 4},
\end{align}
where $f_1$ and $f_2$ are the cumulative covering fractions at $N_{\rm HI}$ in our high-resolution simulations of the low and high-\Tvir~sheets, respectively (Fig \ref{fig:NHI_ResComp_Subregion}, bottom panel). The power law dependence is of course a strong assumption which introduces uncertainty.

We identify the cosmic sheet population in TNG100 using a modified V-web algorithm \citep{Hoffman2012}. We find it much simpler to identify the NTSI-unstable IPM subregions themselves, rather than identifying coherent 4-8 $\rm cMpc/\textit{h}$ sheets. We identify these IPM subregions using the velocity shear tensor to locate planar regions, then require the inflow above and below the planar region converges onto the plane along the corresponding eigenvector of the shear tensor. Appendix \ref{appendix:SheetPops} details IPM-region identification procedure. The algorithm returns all IPM regions in TNG100, along with their associated shock temperatures ($T_{\rm shock}$) and surface areas ($A$). We get the shock temperature of a given IPM region based on the infall velocity perpendicular to the sheet plane, and the area is the sheet volume projected along the inflow eigenvector. 

Approximating $f_{\rm A}$ for all cosmic sheets using only our two simulations is the largest uncertainty here. The full population of cosmic sheets, even at a specific redshift, has a variety of collapse histories. Collapse history affects $f_{\rm A}$ (Fig \ref{fig:AlternativeScenarios}). For example, two sheets with the same infall speed/shock temperature could have different total accumulated sheet masses. To account for this, we use the pressure in the sheet as a proxy for the accumulated mass of the sheet and, thus, the accretion history. We apply a pressure floor ($P/\rm k_b>36~K~ cm^{-3}$), removing any TNG sheets below the pressure of the fragmenting areas in our low \Tvir~sheet simulation. 

Comparing our estimate with observed high-$z$ HI absorber incidence rates \citep{Inoue2014, Crighton2016}, we find that cosmic sheets could account for $\approx1-2\%$ of all LLSs and $\approx10\%$ of all pLLSs at $z=4$. While these numbers are relatively small, we note that absorbers in cosmic sheets should have very low metallicities. Observed ultra low-metallicity ([X/H]$<-3.5$) LLSs and pLLSs make up $\approx2-12\%$ and $\approx5-28\%$, respectively, of the observed $2.2<z<3.6$ population totals \citep{Prochaska2015, Fumagalli2016, Lehner2022}. Thus, our estimate suggests that a substantial fraction of $z=4$ ultra low-metallicity (p)LLSs could have cosmic sheet origins. While our estimates have significant uncertainty, this motivates both improved estimates using more detailed simulations and also potential searches for cosmic sheet association among ultra low-metallicity pLLSs and LLSs.  

We do not constrain how the contribution of sheets to (p)LLSs evolves with cosmic time. At higher redshifts, the mean density and neutral hydrogen column of cosmic sheets increase, but there are fewer large cosmic sheets. These effects both reverse down to lower redshift. How these trends compete to set the abundance of neutral hydrogen absorbers is left for future work.

\subsection{AREPO vs. MP-Gadget} \label{subsec:Arepo_v_MPgadget}

Given the resolution limited nature of the problem, we checked if fragmentation manifests in an equivalent set of simulations run with SPH code MP-Gadget. We found that while we do observe qualitatively similar fragmentation in the SPH suite, shock-broadening and particle smoothing over the cold-core versus hot-virial layer boundary make the SPH simulations more challenging to interpret. Our more detailed results are presented in Appendix \ref{appendix:SPH}.

\section{Summary and Conclusion} \label{sec:Conclusion}

We use high-resolution cosmological simulations to study the morphology and origin of cold gas in cosmic sheets. We simulate two cosmic sheets with the moving-mesh code AREPO, with virial temperatures of $T_{\rm vir}=2\times10^5$ and $8\times 10^5~$K in $4~\rm\cMpc/$$h$ and $8 ~\rm{cMpc}/$$h$ boxes, respectively, spanning gas mass resolutions of $m_{\rm gas}=5\times10^6-9\times10^3~M_{\odot}$. We vary initial conditions and use semi-idealized setups to isolate the mechanism driving gas fragmentation of cosmic sheets. 

Contrary to previous work, we find that the fragmentation of cosmic sheet gas occurs due to nonlinear thin shell instability (NTSI). NTSI grows bending-mode perturbations of the sheet's cold-core. Once this growth saturates, shear-generated vorticity and Kelvin-Helmholtz instabilities (KHI) can fragment the warped cold layer, sometimes into many individual clouds. 

Cold gas morphology in cosmic sheets is poorly converged at typical cosmological simulation resolutions. We find strong enhancements of neutral hydrogen column densities ($N_{\rm HI}>10^{16} \rm~cm^{-2}$). LLS covering fractions experience $\gtrsim2\times$ enhancements between our two highest resolutions. Cosmic sheets may significantly contribute to the populations of very low metallicity pLLSs and LLSs at high redshift. 

Our primary conclusions are as follows:

\begin{itemize}
    \item Resolving cosmic sheets beyond typical large-volume simulations reveals strong kiloparsec-scale IGM fragmentation extending to the resolution limit, consistent with \citet{Mandelker2019, Mandelker2021} (Figs. \ref{fig:FullSheetProjections} \& \ref{fig:SubregionProjections}).
    
    \item This fragmented appearance reflects thickness variations of the low \Tvir~sheet's central cold gas layer, or its disruption into individual clouds in the high \Tvir~sheet (Figs. \ref{fig:PerpendicularSlice_rho} \& \ref{fig:PerpendicularSlice_temp}).

    \item A fourfold increase in spatial resolution raises the pLLS covering fraction by $1.5\times$ in the low \Tvir~sheet, and the LLS covering fraction by $3.5\times$ in the high \Tvir~sheet (Fig. \ref{fig:NHI_ResComp_Subregion}).
                
    \item Semi-idealized sheets, perturbed only on scales below $100 ~\ckpc/h$, fragment similarly to our fiducial, full-$P_{\rm \Lambda CDM}(k)$, sheets (Figs. \ref{fig:Highk_RhoSlice} \& \ref{fig:Highk_TempSlice}). These runs identify NTSI-driven bending as the primary fragmentation mechanism (Fig. \ref{fig:StreamSlices}), with growth times and wavelengths roughly consistent with analytic theory (Section \ref{subsec:NTSItheory}).

    \item Cold-core formation and fragmentation depend on the sheet's collapse history. Variations of our fiducial collapse scenario, e.g. Gaussian vs. tophat overdensity profile or a delayed collapse redshift, yield non-trivial variations (Fig. \ref{fig:AlternativeScenarios}). 

    \item An order-of-magnitude estimate for the fraction of observed HI absorbers in cosmic sheets finds $\mathcal{O}(1\%)$ and $\mathcal{O}(10\%)$ of LLSs and pLLSs, respectively, originate from cosmic sheets at $z=4$. Sheet gas could account for a major fraction of $[\rm X/H<-3.5]$ (p)LLS absorbers, though their precise contribution relative to filaments or halos requires further study (Section \ref{subsec:ObservableLLSs}).

    \item An equivalent set of smoothed particle hydrodynamics simulations shows similar sheet fragmentation, but shock broadening and numerical smoothing between hot- and cold-phase gas hinder interpretation of the fragmentation mechanism (Appendix~\ref{appendix:SPH}).
\end{itemize}

We show that the origin and morphology of cold sheet gas are very sensitive to resolving (sub)kpc spatial scales. Cold gas is important both for its contribution to HI absorption observations and for its potential to fuel star formation if it survives accretion into filaments/halos. We have made initial estimates of the former, but the latter is currently poorly understood at small scales. Future work should explore sheet gas at higher resolutions and across cosmological initial conditions, including studies extending to lower redshifts to track the accretion of IPM cold clouds into filaments and halos.

\begin{acknowledgments}
    We gratefully acknowledge NASA grant 80NSSC24K1220. 
    We gratefully acknowledge the Pacific Postdoctoral Program at the Dark Universe Science Center, University of Washington, during which part of this work was carried out. The Pacific Postdoctoral Program is supported by the Simons Foundation (SFI-MPS-T-Institutes-00012000).
    We acknowledge NSF-ACCESS grants PHY240007 and PHY250179 and the San Diego Supercomputer Center.
    YF is supported by the Tel Aviv University -- Harvard Astronomy Program.
    Anthropic's Claude Code (Opus 5.5 and Fable 5.1) and OpenAI's ChatGPT-5/6 were used for select code and proofreading; all text was written by the authors.
\end{acknowledgments}

\software{numpy \citep{Harris2020}, matplotlib \citep{Hunter2007}, AREPO \citep{Weinberger2020}, MP-Gadget \citep{Feng2018}, yt \citep{Turk2011}, CAMB \citep{Lewis2011}, MP-GenIC \citep{Bird2020}}

\appendix
\twocolumngrid

\section{Fragmentation in SPH} \label{appendix:SPH}
 
Here we discuss the suite of simulations run using the smoothed particle hydrodynamics code MP-Gadget. The MP-Gadget simulations use identical initial conditions and nearly identical physical models to the AREPO ones. The only differences beyond the hydrodynamic algorithm are the SF and self-shielding implementations. In MP-Gadget, we use the built-in `Quick Lyman-$\alpha$-Forest' star-formation prescription with $\delta_{\rm crit}=1000$. Self-shielding is implemented during run-time within MP-Gadget, as opposed to post-processing.

In addition to the fiducial simulation configurations, we run an `ultra-high' resolution run with $N=2\times1024^3$ particles; $m_{\rm gas}\approx \{9,1\}\times10^3 M_{\odot}$ for the high and low \Tvir~sheets, respectively. This higher particle resolution achieves kpc-scale smoothing lengths in the cold gas. 

\begin{figure*}[ht!]
    \epsscale{0.9}
    \plotone{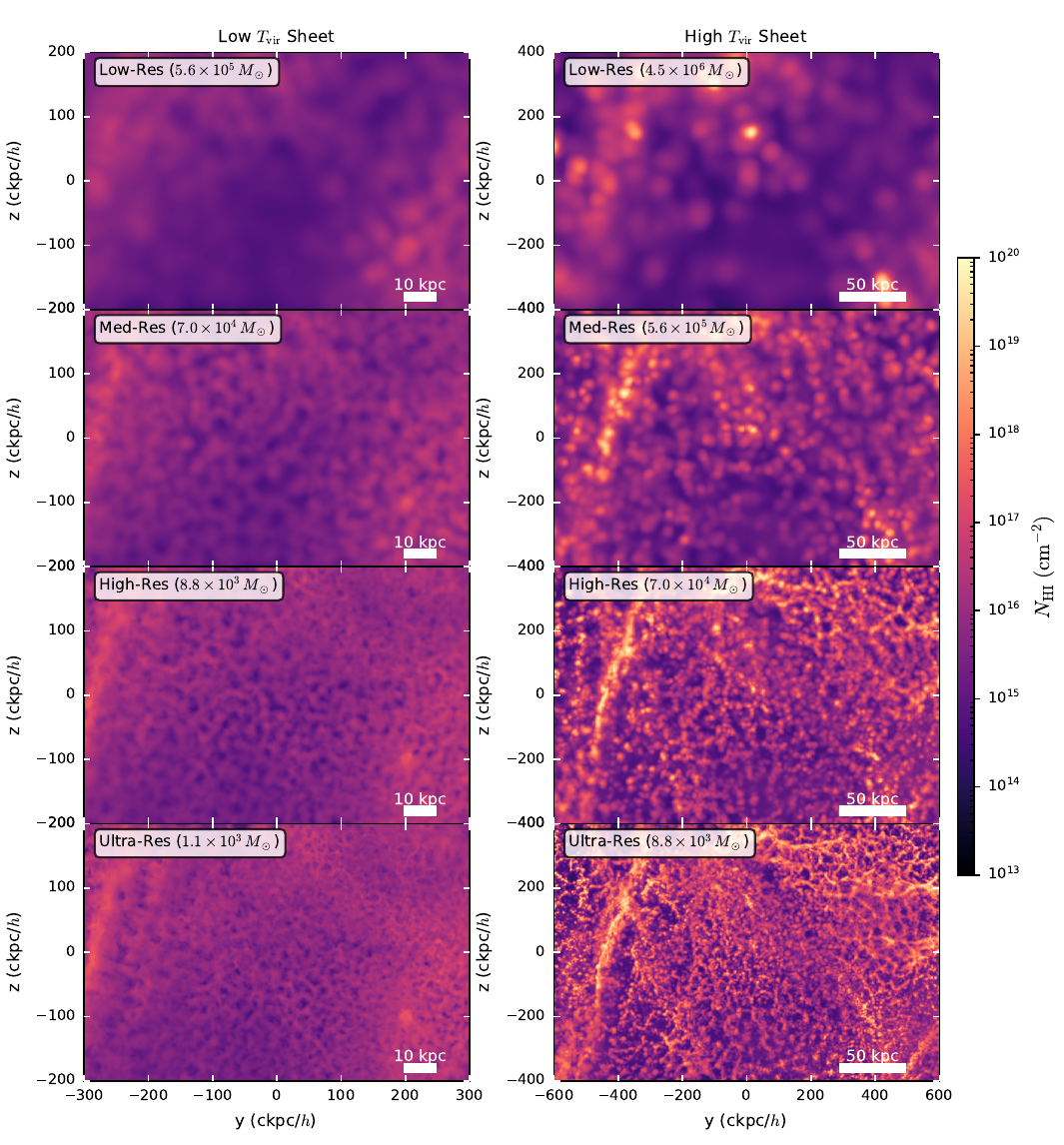}
    \caption{Same as Figure \ref{fig:SubregionProjections}, but for our SPH simulations in MP-Gadget. Rows represent different resolutions ($N_{\rm particles}=2\times \{128, 256, 512, 1024\}^3$), left (right) column is the low (high) \Tvir~sheet. 
    \label{fig:Gadget_Projections}}
\end{figure*}

We find that gas fragmentation does occur in SPH. Figure~\ref{fig:Gadget_Projections} is in the exact same style as Figure~\ref{fig:SubregionProjections}, now showing the MP-Gadget version of the fiducial simulation set. The low \Tvir~sheet looks qualitatively similar to its AREPO counterpart. While showing many similar structures, the morphology of the high \Tvir~sheet appears more fragmented in MP-Gadget. 

For the low \Tvir~sheet, we find $N_{\rm HI}$ distributions in the IPM subregion are similar between AREPO and MP-Gadget; both are unimodal and have similar variance, but MP-Gadget sheets have slightly higher median column density. For the high \Tvir~sheet, however, the $N_{\rm HI}$ distribution is a wide unimodal distribution in MP-Gadget, in contrast to the bimodal distribution in AREPO for the same sheet, presumably caused by smoothing. Going from low to high resolution, the MP-Gadget (AREPO) low \Tvir~cosmic sheet pLLS covering fraction increases from $30\rightarrow 55\%$ ($10\rightarrow 30\%$). The evolution for the LLS covering fraction in the high \Tvir~cosmic sheet is $6\rightarrow 15\%$ ($3\rightarrow 10\%$). The overall trends are similar between the simulations, but with the SPH simulations possessing higher overall neutral hydrogen columns. It is not a new observation that neutral hydrogen predictions differ between moving mesh and SPH \citep{Bird2013}. 

We believe numerical effects play a role in the fragmentation of our SPH simulations, causing some portion of the discrepancies in the $N_{\rm HI}$ distribution and their evolution with resolution between AREPO and MP-Gadget; specifically, shock broadening and over-smoothing of multiphase gas. 

\begin{figure*}[ht!]
    \epsscale{1.15}
    \plotone{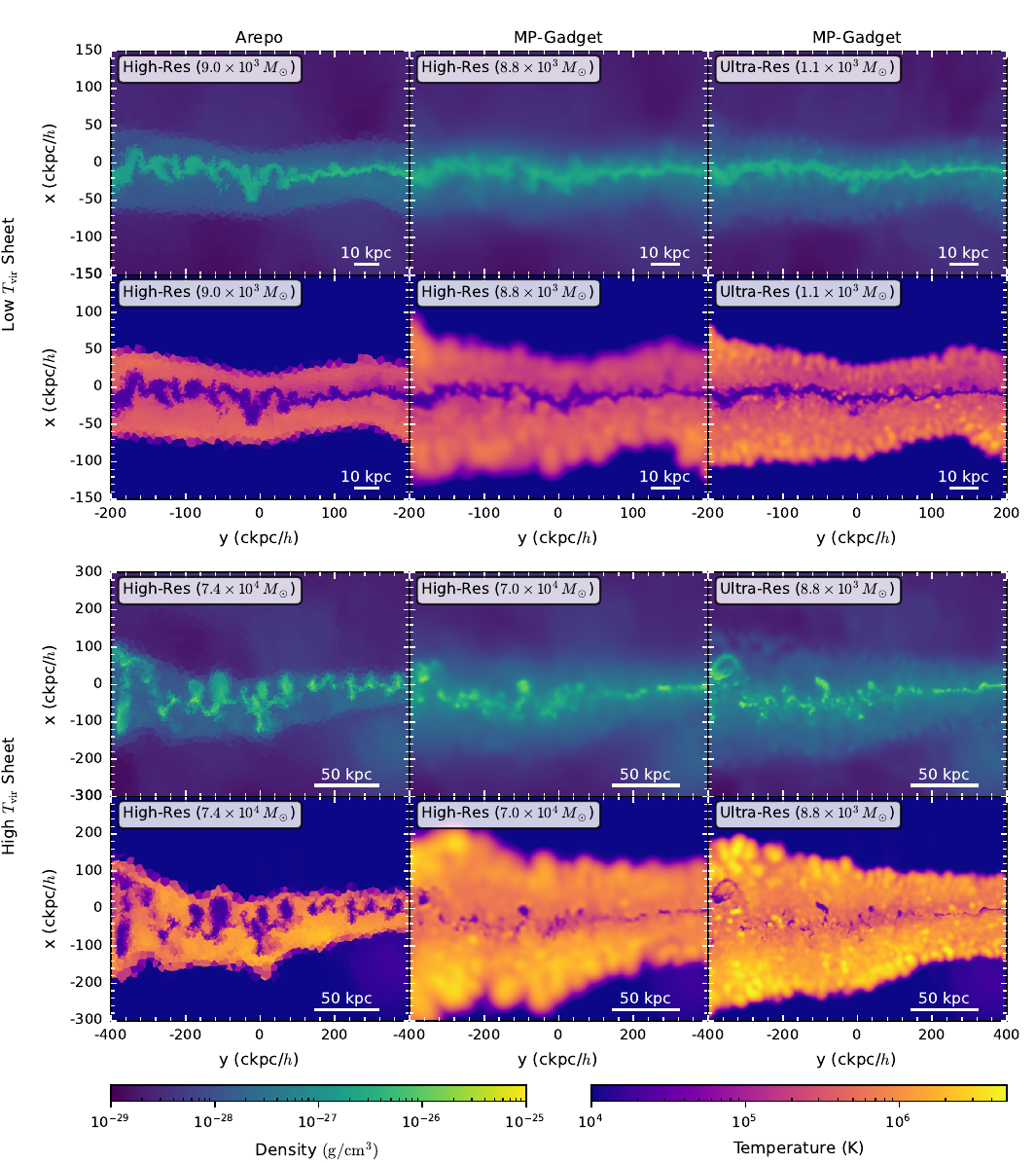}
    \caption{Density and temperature slices of the IPM subregion for the low and high \Tvir~sheets using AREPO and MP-Gadget. The left column is the high-res AREPO sims, the middle column is the high-res MP-Gadget sims, and the right column is the ultra-high-res MP-Gadget sims. All slices are $z=4$.
    \label{fig:Gadget_Slices}}
\end{figure*}

SPH, even with the improved, entropy-conservation formulas (e.g. \citealt{Springel2002}), broadens shocks and destroys sharp gradients on scales around the smoothing length ($l_{\rm smooth}\approx \left[3N_{\rm ngb}m_p/(4\pi \rho\right)]^{1/3}$, $N_{\rm ngb}=27$). Figure \ref{fig:Gadget_Slices} compares the density and temperature slices of the AREPO and MP-Gadget simulations. The hot layer is substantially thicker in SPH. Gas takes a finite time to cross a broadened shock in SPH, causing it to experience over-cooling proportional to the mean cooling rate across the shock and the time taken to cross the shock (i.e., shock width divided by infall speed). Sheet gas shocking to $T=8\times10^5~\rm K$ would normally have a very long, $>1~\rm Gyr$ cooling time. However, as the particle increases in temperature while crossing the broadened shock temperature gradient, it experiences over-cooling via the $10^{4.2}$ and $10^5$~K cooling bumps, overproducing cold gas when $l_{\rm smooth}$ is large. We find the proportion of $z=4$ cold-layer particles with a maximum temperature of $10^{4.2}<\max(T(z))<10^5~\rm K$ decreases by a factor of four from low to high resolution in the SPH simulations. This is the expected effect for a shock which is better resolved in the high resolution simulation. 

The larger challenge in interpreting the physics of cosmic sheet fragmentation in SPH is the smoothing of multiphase gas boundaries. The cold gas layer is of a similar physical thickness as the smoothing length of the hot gas, $h_{\rm cold}\approx l_{\rm smooth}^{\rm hot}\approx\text{a few kpc}$, causing blending of the two distinct gas phases. SPH by construction smooths structures on the scale of the smoothing length. In the case of our cosmic sheets, the hot virial layer can get smoothed through the thin cold layer. In other words, the poor spatial sampling of the cold layer causes nearby hot gas particles to increase the inferred temperature within the cold layer. This causes the blob-like overdensity morphologies observed in both the projections (Fig. \ref{fig:Gadget_Projections}) and slices (Fig. \ref{fig:Gadget_Slices}) of the high \Tvir~sheet. The problem appears to be less catastrophic in the low \Tvir~sheet simulations as the cold layer is thicker there and the resolution higher. Fragmentation is not entirely numerical, as the NTSI/KHI mechanisms should still be present for the SPH sheet, and the low \Tvir~sheets are qualitatively similar. But we could not disentangle the numerical SPH behavior from the underlying fragmentation mechanism.

These numerical effects are less relevant in AREPO because the Voronoi moving mesh algorithm uses a Riemann solver at the faces, which better captures discontinuities, e.g., shocks and multiphase gas. Since AREPO does not smooth over neighboring cells, these simulations have an effective spatial resolution $\approx3\times$ finer for the same number of particles. The tradeoff is the introduction of noise in the gradients without SPH smoothing, and a more complex/expensive hydrodynamic algorithm. With strong shocks, multiphase gas, and low densities, we find that the complex gas physics in cosmic sheets benefits from the Voronoi mesh treatment (despite a large fraction of IGM simulations being done with SPH codes).


\section{Cosmic Sheet Populations in TNG} \label{appendix:SheetPops}

Here we detail how we identified the population of cosmic sheet subregions in TNG100 \citep{Nelson2019} using a modified version of the V-web algorithm \citep{Hoffman2012}. We use the lower resolution TNG100-2 simulation for computational simplicity. We construct a mass-weighted peculiar gas velocity field on a regular grid using nearest-grid-point assignment and smooth this field on a scale $L_{\rm smooth}$ with an FFT tophat kernel.\footnote{Empty voxels are filled with a nearest neighbor value. This is only relevant to void regions, not sheets.} The base V-web algorithm assigns a cosmic-web classification to each voxel at the adopted smoothing scale, but does not partition the sheet-classified volume into individual objects. To measure the surface areas and infall velocities of sheet subregions, we adopt a bottom-up approach: we set $L_{\rm smooth}=0.5~\cMpc/h$ and group adjacent voxels satisfying the selection criteria into connected subregions. $L_{\rm smooth}=0.5~\cMpc/h$ was chosen as the smallest physical scale still comfortably larger than the thickness of a cosmic sheet. 

Because cosmic sheets have irregular geometries and spatially varying infall velocities, a single area and infall speed need not adequately characterize an entire sheet. Our algorithm therefore identifies smaller subregions of approximately planar, converging flow. These IPM subregions provide local analogues of the IPM regions studied in our idealized simulations.

We now describe the criteria each voxel must satisfy to belong to an IPM subregion. We first characterize the local flow using the symmetric peculiar velocity-shear tensor,
\begin{equation}
    \Sigma_{\alpha\beta}
    =
    -\frac{1}{2H(z)}
    \left(
    \frac{\partial v_{{\rm pec},\alpha}}{\partial r_\beta}
    +
    \frac{\partial v_{{\rm pec},\beta}}{\partial r_\alpha}
    \right),
\end{equation}
where $\mathbf{r}$ denotes physical position. Its eigenvalues and eigenvectors satisfy
\begin{equation}
    \boldsymbol{\Sigma}\,\hat{\mathbf{e}}_i
    =
    \lambda_i\,\hat{\mathbf{e}}_i,
    \qquad i=1,2,3,
\end{equation}
with $\lambda_1\geq\lambda_2\geq\lambda_3$. The eigenvalues quantify instantaneous peculiar convergence along the corresponding principal directions, with
\begin{equation}
    \lambda_1+\lambda_2+\lambda_3
    =
    -\frac{\nabla_{\mathbf r}\cdot\mathbf{v}_{\rm pec}}{H(z)}.
\end{equation}

\begin{figure}[ht!]
    \epsscale{1}
    \plotone{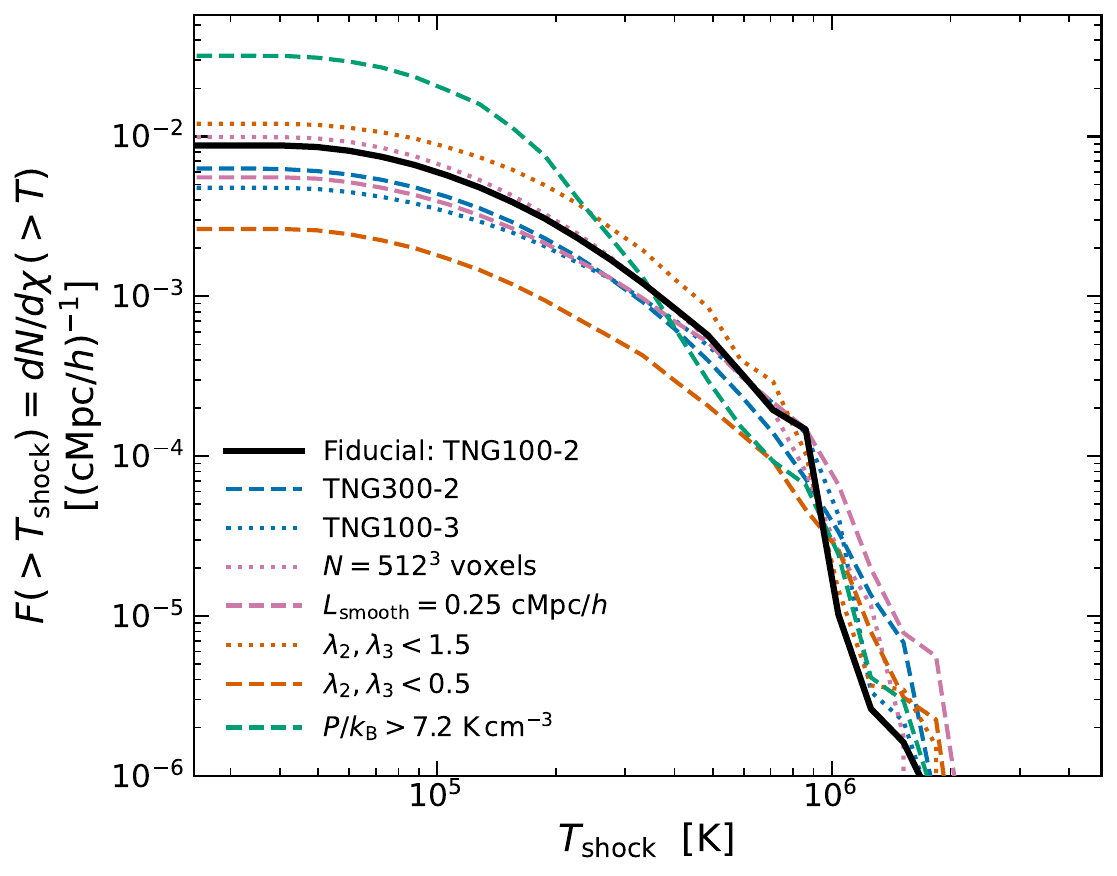}
    \caption{Cumulative LOS incidence of sheet regions vs. shock temperature at $z=4$. The black line denotes the fiducial case used in the main analysis; the fiducial algorithm deposits TNG100-2 onto an $N=1024^3$ voxel grid and applies a smoothing kernel of $L_{\rm smooth}=0.5~ \cMpc/h$, identifying sheet regions using $\lambda_1>1$ and $\lambda_{2,3}<1$, and a pressure floor of $P/k_b>36 ~\rm K~cm^{-3}$. The other lines are all one-parameter ablations of the fiducial model, labeled by the parameter changed. Ablations in legend order: simulation volume, simulation resolution, voxel grid resolution, smoothing kernel size, higher/lower threshold to remove filaments/halos, and the pressure cut threshold.
    \label{fig:TNGsheets_cdf}}
\end{figure}

The standard V-web approach uses an eigenvalue threshold ($\lambda_{i,\rm th}$) above which a voxel is classified as collapsed along the corresponding eigenvector ($\hat{e}_i$). The appropriate threshold depends on the smoothing scale, normalization, and intended application. For example, \citet{Hoffman2012} used $\lambda_{i,\rm th}=0.44$ in their original implementation. We adopt $\lambda_{i,\rm th}=1$, which, applied to sheets ($\lambda_1>1>\lambda_2>\lambda_3$), signifies instantaneous physical contraction along only the first principal direction. Our results are insensitive to variation in the $\lambda_{1,\rm th}$ threshold because additional selection criteria, the minimum infall speed and minimum pressure, impose a partially redundant restriction. We quantify the effect of changing the upper threshold on $\lambda_2$ and $\lambda_3$ (Figure \ref{fig:TNGsheets_cdf}).

The eigenvalues are local quantities, set within $L_{\rm smooth}$, and they identify where the flow is converging, but not the infall speed itself. To measure the infall speed we sample the smoothed peculiar velocity field \textit{non-locally}, above and below each candidate voxel along the sheet normal. The field is evaluated at $\pm r\,\hat{e}_1$ for three probe radii, $r = 0.5,~0.75,~1.0~\cMpc/h$. Writing $\mathbf{v}_{\rm c}$ for the velocity of the candidate voxel, the projected relative velocities above/below the region are
\begin{equation}
    a = \left[\mathbf{v}(+r\hat{e}_1) - \mathbf{v}_{\rm c}\right]\cdot\hat{e}_1,
    \qquad
    b = \left[\mathbf{v}(-r\hat{e}_1) - \mathbf{v}_{\rm c}\right]\cdot\hat{e}_1 ,
\end{equation}
with $a < 0$ and $b > 0$ for gas falling onto the center voxel from either side. We take the infall speed at radius $r$ is simply $v_{\rm in}(r) = \tfrac{1}{2}\left(|a| + |b|\right)$, in which $\mathbf{v}_{\rm c}$ cancels making it independent of the reference frame. The three radii sample the infall velocity profile. We check if the infall speed on each side independently exceeds a floor, $|a| > v_{\rm min}$ and $|b| > v_{\rm min}$. The floor is $v_{\rm min} = 50~\rm km~s^{-1}$, which corresponds to an accretion shock of $T \approx 3\times10^{4}~\K$. We verify that the two sides converge, $a < 0 < b$, and that the relative velocity on each side lies within $30^\circ$ of $\hat{e}_1$, but these two checks are expectedly already satisfied, removing less than 1\% of candidate voxels. If any radius passes all criteria, the voxel is kept and assigned an infall velocity corresponding to the maximum among passing radii. We confirm the peculiar velocities give the best tracer of the shock temperature, rather than the total velocities. The slight underestimate of the infall velocity due to smoothing and measuring at finite distance offsets the slight overestimate from ignoring the Hubble flow. The algorithm, when applied to our fiducial simulations, identified the correct infall velocities, and thus shock temperatures, to within $10\%$.

\begin{figure*}[ht!]
    \epsscale{1}
    \plotone{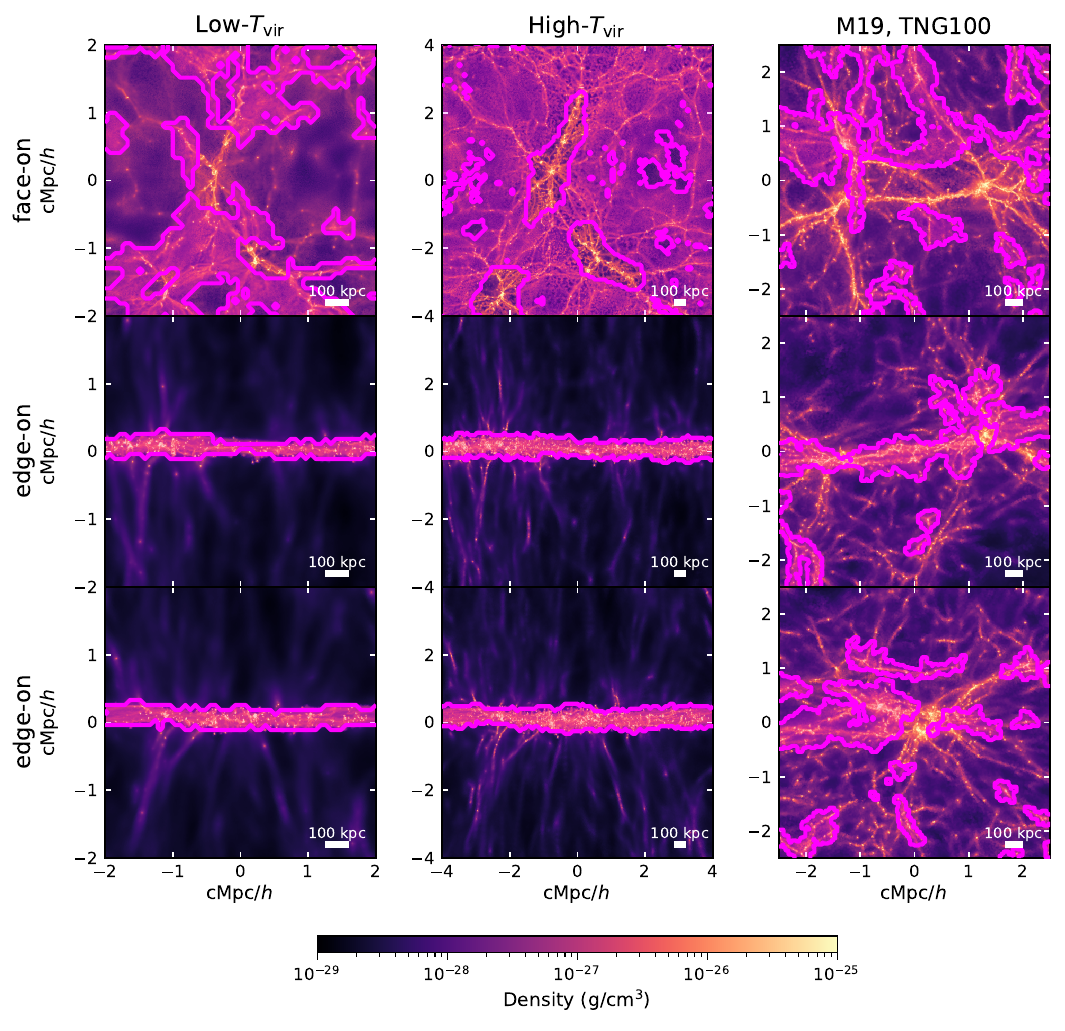}
    \caption{Gas density projections (density-weighted) of the low- and high-\Tvir~sheets (left, middle columns), and the sheet studied by \citet{Mandelker2019} from TNG100-2 (right column). The sheets are shown in the face-on and both edge-on orientations (rows). The pink shading contours show the regions selected by our sheet identification algorithm.
    \label{fig:TNGsheets_proj}}
\end{figure*}

As explained in the main text, our final criterion is a minimum pressure cut: $P>5\times10^{-15}\rm ~erg~cm^{-3}$, i.e. $P/k_{B} > 36~ \rm K~cm^{-3}$. This removes any voxels which are too under-dense for a direct comparison with our idealized sheets. The pressure floor was chosen to cut the non-fragmented regions of our low \Tvir~sheet (Figure \ref{fig:TNGsheets_proj}). 

With a list of converging planar flow voxels, we group together adjacent voxels to identify sheet-regions. We then split voxel groups with misaligned eigenvectors: $\langle| \hat{e}_1 \cdot \bar{e}_{\rm group}|\rangle \leq 0.87$. Subregions of volume less than $0.01 ~(\cMpc/h)^3$ (i.e., $\approx25$ voxels) are removed. We calculate the face-on surface area of each subregion by its projected area along the sheet normal. Each subregion is assigned the median $v_{\rm in}$ of its voxels. We now have the full population of cosmic sheet subregions, each with a specified area and inflow velocity which can be used in Equation \ref{eqn:sheet_abs_incidence}. 

Our specific constraints and criteria were chosen such that they identified the fragmenting regions in our low and high \Tvir~fiducial sheets, as well as the TNG100 cosmic sheet studied in detail by \citet{Mandelker2019}, without identifying the filamentary or node regions. The results of our algorithm as applied to those ground truth cases are shown in Figure \ref{fig:TNGsheets_proj}. The pink shading shows the regions selected as IPM regions. The selected areas of the low \Tvir~sheet (top left) are disjoint because of the pressure cut. The main halos/filaments are removed by the $\lambda_{2,3}<1$ criterion, especially visible in the high \Tvir~sheet (top middle). For all three examples, the $\lambda_1>1$ cut is visible in the edge-on perspective (middle $\&$ bottom rows). Our selections for the \citet{Mandelker2019} sheet are controlled by a mixture of all the algorithm's criteria, emphasizing the more complex morphologies of cosmologically generated sheets. 

The LOS incidence rate of sheet regions as a function of shock temperature is shown in Figure \ref{fig:TNGsheets_cdf}. The figure shows the algorithm run with the fiducial parameters described above (black solid line), as well as ablation runs where a single parameter is varied (colored lines). All runs find that $\mathcal{O}(1\%)$ and $\mathcal{O}(10\%)$ of LLSs and pLLSs, respectively, come from cosmic sheets at $z=4$. The stricter $\lambda_{2,3}<0.5$ cut gave an $\approx2-3\times$ lower results for LLSs and pLLSs (dotted orange line). A $5\times$ lower pressure cut gave an $\approx2-3\times$ higher result (green line). All other ablations had smaller effects. The parameters used were physically motivated, but variations in the algorithm parameters would yield Section \ref{subsec:ObservableLLSs} results lower by a factor of $2-3$.

Note that these constraints alone do not guarantee that NTSI has generated a fragmented cold core layer. For this rough estimate, we do not track whether these conditions are met across cosmic time. In other words, while we confirm these sheet regions are likely unstable to NTSI, we are not applying any growth-time specific conditions. Much of the HI enhancement with resolution comes from resolving the density profile of the sheet. We apply only a first order check on the density via our pressure constraint. Future work should run high-resolution cosmic sheets with a range of densities and collapse times to explore this parameter and its relation to neutral hydrogen enhancement. Our purpose here is to estimate, to first order, whether cosmic sheets could be contributing significantly to HI absorber statistics when properly resolved, but we note the large uncertainties.

\bibliography{references}

\end{document}